\documentclass[journal,10pt]{IEEEtran}
\usepackage{amsthm}
\usepackage{amsmath}
\usepackage{amssymb}
\usepackage{graphicx}
\usepackage{bbm}
\usepackage{algorithm,algpseudocode}
\usepackage{subfig}
\usepackage{cite}
\usepackage{balance}
\usepackage{array}

\makeatletter

\floatstyle{ruled}
\newfloat{algorithm}{tbp}{loa}
\providecommand{\algorithmname}{Algorithm}
\floatname{algorithm}{\protect\algorithmname}

\theoremstyle{plain}
\newtheorem{thm}{\protect\theoremname}
\theoremstyle{definition}
\newtheorem{defn}[thm]{\protect\definitionname}
\theoremstyle{plain}
\newtheorem{lem}[thm]{\protect\lemmaname}

\@ifundefined{showcaptionsetup}{}{%
 \PassOptionsToPackage{caption=false}{subfig}}
\makeatother

\providecommand{\definitionname}{Definition}
\providecommand{\lemmaname}{Lemma}
\providecommand{\theoremname}{Theorem}

\providecommand{\tabularnewline}{\\}
\newcommand{\noun}[1]{\textsc{#1}}

\begin{document}

\title{Fast Adaptation of Activity Sensing\\Policies in Mobile Devices}

\author{Mohammad~Abu~Alsheikh, Dusit~Niyato, Shaowei~Lin, Hwee-Pink~Tan, and Dong In Kim

\thanks{Manuscript received February 29, 2016; revised June 22, 2016 and September 22, 2016; accepted November 07, 2016.}
\thanks{M.~Abu~Alsheikh is with the School of Computer Science and Engineering, Nanyang Technological University, Singapore 639798, and also with the Sense and Sense-abilities Programme, Institute for Infocomm Research, Singapore 138632 (e-mail: stumyhaa@i2r.a-star.edu.sg).\newline
D. Niyato is with the School of Computer Science and Engineering, Nanyang Technological University, Singapore 639798 (e-mail: dniyato@ntu.edu.sg).\newline
S. Lin is with the School of Engineering Systems and Design Pillar, Singapore University of Technology and Design, Singapore 487372 (e-mail: shaowei\_lin@sutd.edu.sg).\newline
H.-P. Tan is with the School of Information Systems, Singapore Management University, Singapore 188065 (e-mail: hptan@smu.edu.sg).\newline
D. I. Kim is with School of Information and Communication Engineering, Sungkyunkwan University, Korea  440-746 (e-mail: dikim@skku.ac.kr).}
}
{}

\maketitle
\begin{abstract}
With the proliferation of sensors, such as accelerometers, in mobile devices, activity and motion tracking has become a viable technology to understand and create an engaging user experience. This paper proposes a fast adaptation and learning scheme of activity tracking policies when user statistics are unknown a priori, varying with time, and inconsistent for different users. In our stochastic optimization, user activities are required to be synchronized with a backend under a cellular data limit to avoid overcharges from cellular operators. The mobile device is charged intermittently using wireless or wired charging for receiving the required energy for transmission and sensing operations. Firstly, we propose an activity tracking policy by formulating a stochastic optimization as a constrained Markov decision process (CMDP). Secondly, we prove that the optimal policy of the CMDP has a threshold structure using a Lagrangian relaxation approach and the submodularity concept. We accordingly present a fast Q-learning algorithm by considering the policy structure to improve the convergence speed over that of conventional Q-learning. Finally, simulation examples are presented to support the theoretical findings of this paper.
\end{abstract}

\begin{IEEEkeywords}
Activity tracking, fast adaptation, Internet of Things, Markov decision processes, wireless charging.
\end{IEEEkeywords}
\IEEEpeerreviewmaketitle

\section{Introduction}\label{sec:introduction}

Activity tracking promises to revolutionize mobile user experience and helps in understanding the big data of today's world~\cite{lara2013survey,perera2014context}. Specifically, activity and motion data is required in many applications such as home security and automation, healthcare systems, contextual advertising, and smart vehicle technologies. Using mobile devices, such as mobile phones and Internet of Things~(IoT) gadgets, for activity tracking has many benefits over conventional wearable sensor and body networks in terms of reachability, flexibility, and financial cost. Firstly, the mobile phone market has been rapidly scaling with more than $63\%$ international penetration rate in 2015 and 4.7 billion unique mobile subscribers~\cite{page2016mobile}. Secondly, modern mobile devices are equipped with high-quality built-in sensors that can measure various physical quantities such as orientation, motion, ambient light, and location. Thirdly, mobile devices support efficient data transmission using cellular networks which facilitates backend integration and data synchronization. Fourthly, application stores, such as Google Play, enable the reach of a huge customer base for mobile crowdsensing in activity-aware systems and support software and patch upgrades.

Continuous activity and motion tracking is being deterred by the energy and monetary cost of mobile sensing. Firstly, mobile devices are battery-powered, and they deplete their energy within a few hours when in-device sensors operate continuously~\cite{wang2009framework,lu2010jigsaw,wang2010markov,yurur2014energy,chon2014adaptive}. Wireless charging is gaining an increasing attention from hardware manufacturing companies as a seamless recharging method of mobile devices. The authors in~\cite{jadidian2014magnetic} showed that a mobile device can be remotely charged using magnetic resonance coupling while being in user's pocket. Nonetheless, wireless charging is intermittent due to mobility and is not available at all locations. Secondly, cellular data plans are generally expensive, and continuous activity tracking can cause significant bill overcharges by cellular operators for data transmission and synchronization with a backend. Data synchronization is typically required for an up-to-date tracking of user activities and motion over time, and hence provide customized mobile services accordingly.

To address these issues, this paper proposes an adaptive activity tracking policy for mobile devices, and considers the intermittent (wired and wireless) charging and cellular data usage. The mobile sensing optimization is designed to minimize the detection error of user activities subject to a data usage limit. The main contributions and results of this paper are summarized as follows:
\begin{itemize}
\item In Section~\ref{sec:mobile_sensing}, the activity tracking problem is formulated as a stochastic optimization using constrained Markov decision processes (CMDPs). A CMDP model~\cite{altman1999constrained} is a variant of Markov decision processes (MDPs) for stochastic optimization subject to a constraint on problem variables and feasible solutions. The temporal correlation of user activities is modeled as a discrete-time Markov chain (DTMC), and the CMDP tracking policy minimizes the detection error of user activities subject to a predefined data usage constraint. 
\item Using a Lagrangian relaxation approach~\cite{beutler1985optimal,sennott1993constrained}, we relax the CMDP formulation to an unconstrained MDP  as discussed in Section~\ref{sec:unconstrained_MDP}. Then, the optimal tracking policy is found using conventional solution methods such as the value iteration algorithm. A Lagrange multiplier in the unconstrained optimality equation is found recursively to capture the data usage constraint.
\item Based on the unconstrained MDP, the threshold structure of the policy is proved in Section~\ref{sec:structure_analysis}. The MDP activity tracking policy is shown to be monotonically non-decreasing in the battery level, and hence the CMDP policy can be represented as a mixture of two threshold MDP policies. Accordingly, fast adaptation of Q-learning can be achieved based on the proved threshold structure of the activity tracking policy. 
\end{itemize}

The remainder of this paper is organized as follows. Section~\ref{sec:related_work} reviews related works in the literature. Section~\ref{sec:mobile_sensing} presents a CMDP activity tracking policy for mobile devices. Then, the CMDP activity tracking policy is transformed to an unconstrained MDP using a Lagrange-based method as presented in Section~\ref{sec:unconstrained_MDP}. Based on the unconstrained MDP problem, the threshold structure of the activity tracking policy is proved in Section~\ref{sec:structure_analysis}, and a threshold Q-learning method that leverages the structure of the activity tracking policy is also discussed. The performance evaluation is presented in Section~\ref{sec:numerical_validation}. Finally, the paper is concluded in Section~\ref{sec:conclusions}.

\section{Related Work}\label{sec:related_work}
In this section, we first review related works on the applications of activity tracking systems. Then, we discuss the wireless charging technologies available for mobile devices. Finally, we review related works on mobile sensing optimization.

\subsection{Activity Tracking Using Mobile Devices}
Mobile devices can be programmed to sense and adapt to the physical environment. The authors in~\cite{miluzzo2007cenceme} presented an algorithm for detecting human contexts, e.g., activities, disposition, and habits, which can be integrated with social networking services. A security method that uses human gestures for continuous authentication was proposed in~\cite{feng2014tips}. In~\cite{milovsevic2011applications}, a mobile sensing application in healthcare systems was discussed. The application monitors human physical activities, e.g., heart activity, to generate continuous feedback on health and behavior conditions. The authors in~\cite{qin2014tagsense} proposed a method that infers user activities for automatic image tagging. Specifically, the rich tags include information about user activities and location, and surrounding ambient light and sound.

\subsection{Wireless Chargers for Mobile Devices}
Wireless charging of mobile devices has seen great advancements in the last few years. This enables the remote charging of devices at a distance of a few meters, i.e., the mobile device is not required to be on the wireless charging pad as in the old technology. The authors in~\cite{jadidian2014magnetic} proposed a wireless charging system, called ``MagMIMO'', for mobile devices based on the technology of magnetic resonance coupling. Similar to beamforming in multiple-input-multiple-output (MIMO) antennas, the proposed system embeds multiple coils in the power charger, and hence forms the magnetic field as a beam focused towards the mobile device. MagMIMO enables effective charging of one mobile device at a distance of $0.4$m from the charger. The authors in~\cite{shi2015wireless} introduced ``MultiSpot'', a wireless charging system based on magnetic resonance which can charge up to $6$ mobile devices simultaneously. The effective charging distance is $0.5$m. MultiSpot uses multiple coils in the wireless charger to beam the charging signal towards the mobile devices. ``Wattup''~\cite{wattup2016} is a wireless charger of mobile devices that uses radio frequency (RF) radiation with an effective charging distance of $15$ feet ($4.57$m). The mobile devices are first located using low-energy Bluetooth signals. After the successful localization, an RF signal, similar to the WiFi signal, is sent in the direction of the mobile devices. Wattup comes with a controlling software to select the devices to be charged. Cota~\cite{cota2016} is another product that is based on the RF radiation technology. The effective charging distance is $10$m.

These recent advancements have encouraged many companies to support wireless chargers in their products and stores. For example, IKEA, the international furniture retailer, has established a new production line that embeds wireless chargers in the furniture and home accessories~\cite{ikea2016}. Starbucks has started to install wireless chargers in some of its coffee shops worldwide~\cite{starbucks2016}.

\subsection{Mobile Sensing Optimization}
Optimal mobile sensing of user activities is typically designed to maximize the detection accuracy under a resource constraint. For example, the authors in~\cite{wang2009framework} used the knowledge about user's motion, location, and surrounding environment to manage sensor activation for detecting various activities. Specifically, the sensor activation is semi-automated and is based on manual settings and an apriori distribution. A related MDP-based method was also presented in~\cite{chon2014smartdc} to continuously model the user mobility. The continuous sensing is avoided by exploring the location information. The authors in~\cite{lu2010jigsaw} presented an algorithm that uses accelerometer, microphone and GPS sensors to detect human activities and balances the detection performance and energy consumption. In~\cite{wang2010markov}, the mobile sensing problem was formulated as a CMDP. The design objective is to maximize the detection accuracy under a given energy constraint. Similarly, the authors in~\cite{yurur2014energy} proposed mobile sensing algorithms for accelerometer-based systems using CMDP and partially observable MDP models. The user behavior is assumed to be time-varying which is captured by statistical methods, e.g., the entropy-production rate. In~\cite{chon2014adaptive}, the sensor activation of a mobile tracking system was formulated using a hidden Markov model. The mobility pattern, residual energy, and cellular connection are evoked in predicting a schedule for sensor activation, e.g., a GPS sampling schedule.

This paper substantially differs from existing works in terms of the problem formulation, optimization objectives and constraints, and results. Existing works on activity sensing and tracking in the literature do not consider the user adaptation of tracking policies. Therefore, learning a policy for a particular user with conventional methods requires a large number of iterations which is expensive in mobile devices. This paper has clear novelty in providing fast user adaptation of activity tracking policies. In particular, the theoretical analysis employs a Lagrange relaxation approach along with the concept of submodularity to prove that the CMDP policy is a randomized mixture of two threshold MDP policies that are monotonically non-decreasing in the battery level. Our threshold analysis (a)~enables fast online policy learning, e.g., using Q-learning, by substantially reducing the search space of the optimal activity tracking policy, and (b)~curtails the storage space of the policy. 

\section{Problem Formulation and Optimization}\label{sec:mobile_sensing}

\begin{table}
\caption{List of frequently used symbols throughout the paper.\label{tab:list_symbols}}
\renewcommand{\arraystretch}{1.2}
\begin{tabular}{|c|>{\centering}p{0.62\columnwidth}|}
\hline 
\textbf{\noun{Symbol}} & \textbf{\noun{Definition}}\tabularnewline
\hline 
\hline 
$\mathcal{U}$ & User activity state space\tabularnewline
\hline 
$\mathcal{E}$ & Energy charging state space\tabularnewline
\hline 
$\mathcal{B}$ & Battery level state space\tabularnewline
\hline 
$\psi^{n}=(u^{n},e^{n},b^{n})$ & System state at time $n$ containing the user activity $u^{n}$, number of acquired energy units~$e^{n}$, and battery level
$b^{n}$\tabularnewline
\hline 
$\Delta=\{ \delta_{0}, \delta_{1} \}$ & Action space defining the sleep~$\delta_{0}$ and active~$\delta_{1}$ modes\tabularnewline
\hline 
$c(\cdot)$ & Detection error function of the user activities\tabularnewline
\hline 
$g(\cdot)$ & Probability of connectivity to an access network\tabularnewline
\hline 
$\mathbb{P}\left(\psi^{n+1}|\psi^{n},\delta^{n}\right)$ & Transition probability from state $\psi^{n}$ to state $\psi^{n+1}$
after taking action $\delta^{n}\in\Delta$ at time $n$\tabularnewline
\hline 
$d(\cdot)$ & Data usage function for data synchronization with a backend\tabularnewline
\hline 
$\pi$&Activity sensing policy defining the sensing action at each state\tabularnewline
\hline 
$\ensuremath{\mathcal{J}\left(\cdot\right)}$ & Average detection error under policy $\pi$\tabularnewline
\hline 
$\ensuremath{\mathcal{D}\left(\cdot\right)}$ & Average data usage under policy $\pi$\tabularnewline
\hline 
$\lambda$ & Lagrange multiplier\tabularnewline
\hline 
$\beta$ & Discount factor in the unconstrained MDP formulation\tabularnewline
\hline 
$\bar{b}$ & Average battery level\tabularnewline
\hline 
$\rho$ & Probability of successful data synchronization\tabularnewline
\hline 
$\tau$ & Probability of battery overflow\tabularnewline
\hline 
\end{tabular}
\end{table}

\subsection{Overview}

\begin{figure}
\begin{centering}
\includegraphics[width=0.85\columnwidth,trim=2cm 1.5cm 2cm 1cm]{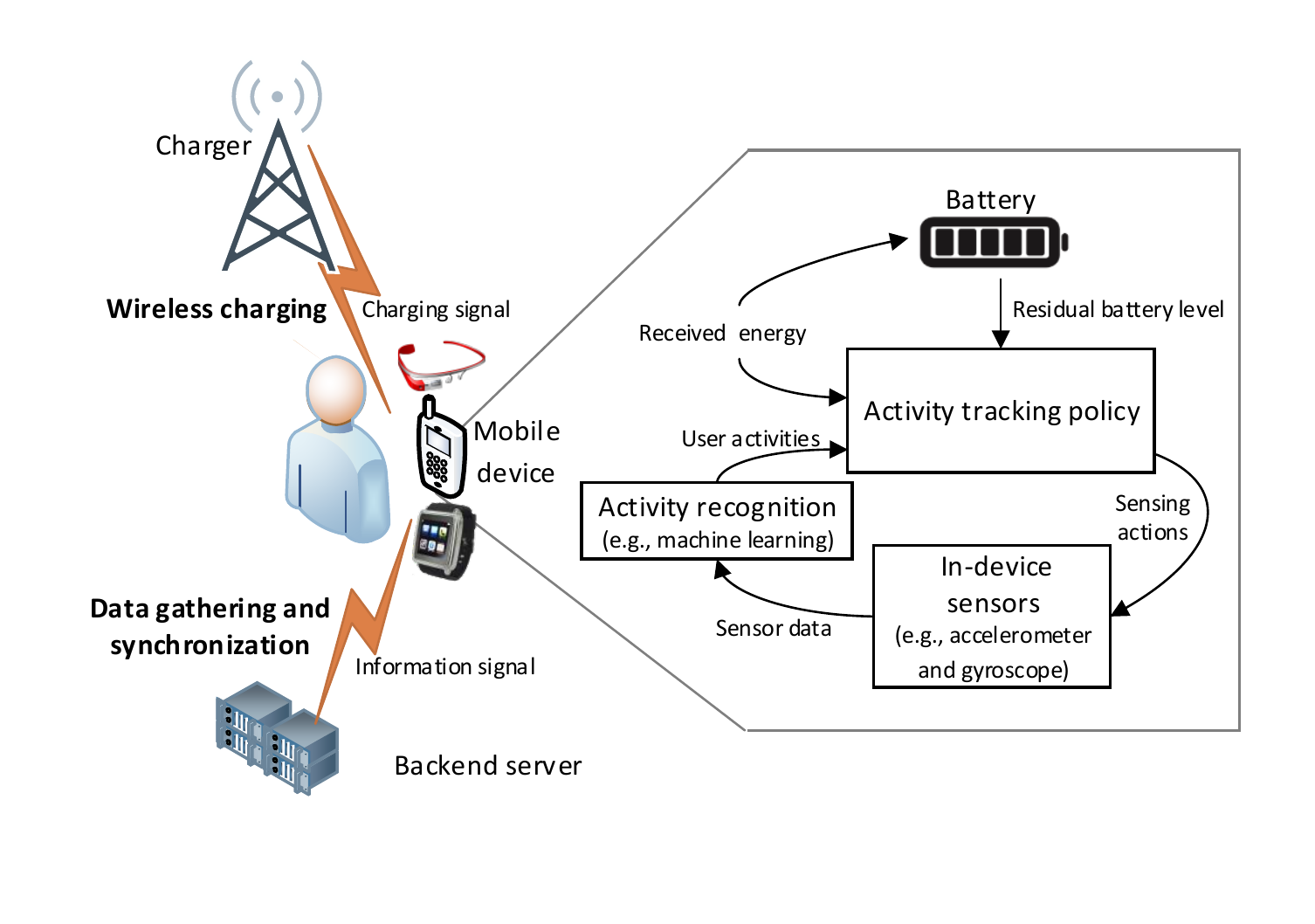}
\par\end{centering}
\caption{System model including data synchronization and intermittent charging, e.g., wireless charging.\label{fig:system_model}}
\end{figure}

Activity and motion tracking is becoming an integral functionality in modern mobile platforms such as Google Fit for Android\footnote{https://fit.google.com/}, Apple Health for iOS\footnote{http://www.apple.com/sg/ios/health/}, and Motion Data 2.0 for Windows Phone\footnote{http://windows.microsoft.com/en-gb/windows-10/motion-data-privacy-faq}. The detected user activities can be shared with all applications installed in a mobile device to provide interactive user experience. A modern example of activity-aware schemes is targeted advertising for dynamically delivering an advertisement based on user's activities and disposition which increases the revenue of both publishers and advertisers~\cite{sala2007exploration}. Figure~\ref{fig:system_model} shows the system model as considered in this paper. The mobile sensing is systematically managed based on the user activity, battery level, and battery charging state of a mobile device. These parameters are used for taking decisions on optimal working modes. There are two working modes: (i)~an active mode during which battery charging, data synchronization, and activity sensing can be performed, and (ii)~a sleep mode during which only battery charging is performed. The activity detector maps time series data into the most probable user activity using supervised machine learning techniques. This mapping process is beyond the scope of this paper. Nonetheless, we refer interested readers to~\cite{parkka2006activity,alsheikh2016mobile} for some pertinent results. The list of symbols used in this paper are summarized in Table~\ref{tab:list_symbols}.

In this section, the mobile sensing problem is formulated as a finite state Markov model. To summarize, this section describes the following steps.
\begin{itemize}
\item Similar to previous works in~\cite{ziebart2008navigate,wang2010markov,yurur2013adaptive,chon2014smartdc}, user activities are assumed to evolve as a discrete-time Markov chain (DTMC). Then, the activity and motion tracking is formulated as an infinite horizon CMDP model with a single constraint. The CMDP consists of: decision epochs, system states, actions, transition probabilities, detection error and data usage functions, and a cellular data limit. The objective is minimizing the detection error of user activities subject to a cellular data limit. 
\item It follows from the results in~\cite{altman1999constrained,heyman2003stochastic} that the CMDP sensing problem presented as a linear program can be solved in polynomial time to obtain a randomized, stationary, and optimal activity tracking policy.
\end{itemize}

\subsection{System Model}

In what follows, an activity-aware mobile device is assumed to operate under a discrete time fashion with decision epochs denoted by $\mathcal{N}=\left\{ 1,2,\ldots,N\right\} $, where $N$ is the sequence termination time. At each decision epoch $n\in\mathcal{N}$, the system selects a mobile sensing action based on its current system state and then transits to a new state. The system state space $\Psi$ of the mobile sensing problem is defined as follows:
\begin{multline}
\Psi=\bigg\{\left(\mathcal{U},\mathcal{E},\mathcal{B}\right),\\
\mathcal{U}\in\left\{ 0,\ldots,U\right\};\mathcal{E}\in\left\{ 0,1\right\};\mathcal{B}\in\left\{ 0,\ldots,B\right\} \bigg\},
\end{multline}
where $\mathcal{U},\mathcal{E}$, and $\mathcal{B}$ represent the user activity state, energy charging state, and battery level of the mobile device, respectively. $U$ is the maximum number of supported user activities, and $B$ is the maximum capacity of the battery of energy units. Consequently, state $\psi^{n}\in\Psi$ at time $n\in\mathcal{N}$ is defined using a 3-tuble as $\psi^{n}=(u^{n},e^{n},b^{n})$ which includes the current user activity $u^{n}$, the number of newly acquired energy units $e^{n}$ ($e^{n}\in\left\{ 0,1\right\} $), and the battery level $b^{n}$ at that decision epoch. Similar to previous works in~\cite{ziebart2008navigate,wang2010markov,yurur2013adaptive,chon2014smartdc}, the user activity states are assumed to evolve as a Markov chain with transitions that are stochastically involved. Under discrete time model, this assumption is typical as user activities have short memory.

The action space $\Delta = \{ \delta_{0}, \delta_{1} \}$ includes two actions as follows:
\begin{equation}
\begin{cases}
\delta_{0}=0, & \textrm{switch to the sleep mode},\\
\delta_{1}=1, & \textrm{switch to the active mode.}
\end{cases}
\end{equation}
During the active mode, the mobile device can measure samples using its in-device sensors, e.g., an accelerometer, gyroscope, digital compass, microphone, and GPS. Additionally, the data synchronization can only be performed during the active mode. The mobile device is assumed to consume one unit of energy during the active mode while no energy is depleted in the sleep mode. It is important to note that these modes of operations are restricted to the activity-aware system and are separated from other applications running on the device. The detection error function of user activities $c(\cdot)$ is defined as
\begin{equation}
c(\cdot):\Psi\times\Delta\rightarrow\mathbb{R}_{+}.\label{eq:cost_function}
\end{equation}
$c(\cdot)$ is a decreasing function on the user activities which is defined by considering the detection error of each activity by machine learning algorithms, i.e., $u_{0}\in\mathcal{U}$ has the highest detection error. For example, the authors in~\cite{parkka2006activity} used decision trees and supervised neural networks to classify daily human activities with varying detection errors of $3.0-42.0\%$ and $4.0-78.0\%$, respectively. In~\cite{alsheikh2016mobile}, a deep learning model is designed to detect human activities from crowdsensing data which scores $3.0-47.0\%$ of varying detection errors. Clearly, user activities cannot be detected during the sleep mode $\delta_{0}$ such that $c(\psi,\delta_{0})=1.0$.

Another practical advantage of the proposed model is its consideration of the mobile device connectivity to a backend. This is important as the data connectivity depends on the available access networks in the area~\cite{doufexi2003hotspot}. The probability of having wireless connection to an access network at each system state $g(\cdot)$ is defined as 
\begin{equation}
g(\cdot):\Psi\times\Delta\rightarrow[0,1],\label{eq:coverage_probability}
\end{equation}
where $\times$ is the Cartesian product. $g(\cdot)$ can be defined based on the user activity, battery level, and the actions of the activity-aware mobile device. The connectivity probability to the activity-aware backend is zero when the mobile device is switched to the sleep mode, i.e., $g(\psi,\delta_{0})=0$, as no data transmission is allowed.

For simplicity, the battery charging probability is assumed to be sampled from a Bernoulli distribution with a success probability of $\mathbb{P}(e=1)$. This charging probability can be construed as the probability that a mobile device is able to receive energy from a wireless charger. Nonetheless, other more complex distributions can be adopted without affecting the problem formulation. The battery capacity is finite, takes integer values only, and follows Lindley equation \cite{gross2008fundamentals} which is given as follows:
\begin{equation}
b^{n+1}=\min\left(\left[b^{n}-\delta^{n}\right]^{+}+e^{n},B\right),
\end{equation}
where $\left[\cdot\right]^{+}$ is defined as $\left[z\right]^{+}=z$ when $z>0$, and it returns $0$ otherwise. During one time epoch of the battery charging, one energy unit is added to the battery of the mobile device unless the battery is full, i.e., the maximum capacity of a battery is finite and is replenished by wired or wireless charging. The mobile device consumes one unit of energy during a time epoch of active mode. This energy is used for both activity sensing, processing, and synchronization. We remark that our optimization model can be extended straightforwardly for arbitrary number of units of energy consumption and depletion, e.g., the received energy can change based on the distance between the wireless charger and the mobile device~\cite{jadidian2014magnetic}.

With the above setup, the transition probability $\mathbb{P}\left(\psi^{n+1}|\psi^{n},\delta^{n}\right)$ from state $\psi^{n}=(u^{\text{n}},e^{n},b^{n})\in\Psi$ at time $n\in\mathcal{N}$ to state $\psi^{n+1}=(u^{n+1},e^{n+1},b^{n+1})\in\Psi$ at time $n+1\in\mathcal{N}$ after taking action $\delta^{n}\in\Delta$ at time $n\in\mathcal{N}$ is found as follows:
\begin{multline}
\mathbb{P}\left(\psi^{n+1}|\psi^{n},\delta^{n}\right)= \mathbb{P}(u^{n+1}|u^{n})\mathbb{P}(e^{n+1})\\
\times\bigg[\mathbbm{1}(b^{n+1}=b^{n}+e^{n}-\delta^{n})g(\psi^{n},\delta^{n})	\\
+\mathbbm{1}(b^{n+1}=b^{n}+e^{n})\left(1-g(\psi^{n},\delta^{n})\right)\bigg],
\label{eq:transition}
\end{multline}
where $\mathbbm{1}(\cdot)$ is an indicator function which is used to maintain consistent battery levels over time due to energy consumption and depletion. $\mathbb{P}(u^{n+1}|u^{n})$ is the probability of transiting between user activities. Furthermore, $\mathbb{P}(e^{n+1})$ is the probability of the mobile device to be in the charging mode.


Recall that the system is also assumed to be pertaining under a data usage constraint $D$. This constraint is important to avoid overcharges by cellular operators for data synchronization to a backend. Therefore, we define $d(\cdot)$ as the data usage function which returns a non-negative value based on the taken actions $d(\cdot):\Delta\rightarrow\mathbb{R}_{+}$. Mathematically, $d(\cdot)$ is defined as follows:
\begin{equation}
d(\psi,\delta)=\begin{cases}
d(\psi,\delta_{1}), & b>0\textrm{ and }\delta=1,\\
0, & \text{otherwise}.
\end{cases}\label{eq:data_usage_function}
\end{equation}
If the battery is not empty, i.e., $b>0$, and the active action $\delta_{1}$ is selected, activity data packets are generated and transmitted to the backend. Here, there is an important connection between the cellular data limit $D\in\mathbb{R}_{+}$ and the data generated during the active mode $d(\psi,\delta_{1})$. Specifically, a mobile device transmits activity data to a backend with probability $\xi$ of the total epochs such that 
\begin{equation}
\xi=\frac{D}{d(\psi=[u,e,b],\delta_{1})}.\label{eq:coverage_percentage}
\end{equation}
For example, the mobile device transmits activity data during one fourth of its total decision  epochs when $d(\psi=[u,e,b],\delta_{1})=1$ and $D=0.25$.

Before proceeding further with the problem solution, we define a \emph{decision rule} $\pi{}_{n}$ at time epoch $n$ as a mapping between the current system state and the optimal action $\pi_{n}:\Psi\rightarrow\Delta$. A \emph{policy} $\pi\doteq\left(\pi_{1},\pi_{2},\ldots,\pi_{N}\right)$ is a sequence composition of optimal decision rules through all decision epochs. A policy is called as a \emph{stationary policy} if its decision rules are not changing over time. A major objective of the activity tracking policy is to minimize the overall error of monitoring the user activities by selecting optimal actions $\delta\in\Delta$ over time. Therefore, we denote the optimal, stationary tracking policy as $\pi^{*}\left(\psi,\delta\right)$ which maps state $\psi\in\Psi$ and action $\delta\in\Delta$.

\subsection{Optimal Activity Tracking Policy}

As our system design imposes a constraint on the data usage, the mobile sensing problem is formulated as a CMDP which is expressed as follows:
\begin{eqnarray}
\min_{\pi} & \mathcal{J}\left(\pi\right)=\lim\limits _{N\rightarrow\infty}\sup\frac{1}{N}\sum\limits _{n=1}^{N}\mathbb{E}(c(\psi^{n},\delta^{n}))	,\label{eq:cmdp_cost}\\
\textrm{s.t.} & \mathcal{D}\left(\pi\right)=\lim\limits _{N\rightarrow\infty}\sup\frac{1}{N}\sum\limits _{n=1}^{N}\mathbb{E}(d(\psi^{n},\delta^{n}))	\leq D,\label{eq:eq:cmdp_constraint}
\end{eqnarray}
where $\psi^{n}\in\Psi$ and $\delta^{n}\in\Delta$ are the state and action at time $n$, respectively. $\mathcal{J}\left(\cdot\right)$ and $\mathcal{D}\left(\cdot\right)$ are the average detection error and data usage under policy $\pi$, respectively. $\mathbb{E}(\cdot)$ is the expectation function, and $\pi^{*}(\psi,\delta)$ is an optimal activity tracking policy that defines the probability of taking action $\delta$ at state $\psi$. $D\in\mathbb{R}_{+}$ is the  data usage limit for the transmission of user activities to the backend. The objective function in (\ref{eq:cmdp_cost}) minimizes the detection error subject to a data usage limit given by~(\ref{eq:eq:cmdp_constraint}).

It has been shown in~\cite{heyman2003stochastic,altman1999constrained} that a CMDP model can be solved using linear programming (LP) in polynomial time. Let $\phi(\psi,\delta)$ denote the stationary probability of state $\psi$ and action $\delta$. The mobile sensing problem can be formulated as follows:
\begin{eqnarray}
\min_{\phi(\psi,\delta)} && \sum\limits _{\psi\in\Psi}\sum\limits _{\delta\in\Delta}\phi(\psi,\delta)c(\psi,\delta),\label{eq:cmdp_1}\\
\textrm{s.t.} & & \sum\limits _{\psi\in\Psi}\sum\limits _{\delta\in\Delta}\phi(\psi,\delta)d(\psi,\delta)\leq D,\label{eq:cmdp_2}\\
 & & \sum\limits _{\delta\in\Delta}\phi(\psi',\delta)=\sum_{\psi\in\Psi}\limits\sum\limits _{\delta\in\Delta}\phi(\psi,\delta)\mathbb{P}(\psi'|\psi,\delta),\label{eq:cmdp_3}\\
  & & \sum\limits _{\psi\in\Psi}\sum\limits _{\delta\in\Delta}\phi(\psi,\delta)=1,\phi(\psi,\delta)\geq0,\label{eq:cmdp_4}
\end{eqnarray}
where $\psi'\in\Psi$. The solution of this problem is the optimal stationary probability $\phi^{*}(\psi,\delta)$. The objective function in (\ref{eq:cmdp_1}) minimizes the activity detection error. The constraint in (\ref{eq:cmdp_2}) maintains the cellular data usage below a target level $D$. Then, the constraint in (\ref{eq:cmdp_3}) ensures the ergodic transition between system states. The constraints in (\ref{eq:cmdp_4}) assert on stationary probability requirements. Solving (\ref{eq:cmdp_1})-(\ref{eq:cmdp_4}) using an LP solver gives the optimal stationary probability $\phi^{*}(\psi,\delta)$. The optimal policy is then found for each state and action pair as $\pi_{\textrm{CMDP}}^{*}(\psi,\delta)=\frac{\phi^{*}(\psi,\delta)}{\sum_{\delta'\in\Delta}\phi^{*}(\psi,\delta')}$. $\pi_{\textrm{CMDP}}^{*}$ is a randomized, stationary policy~\cite{altman1999constrained}, i.e., $\pi_{\textrm{CMDP}}^{*}$ is randomized over available actions and does not change over time.

\subsection{Threshold Activity Tracking Policy}
\begin{figure}
\begin{centering}
\includegraphics[width=0.9\columnwidth,trim=1cm 1cm 1cm 0.5cm]{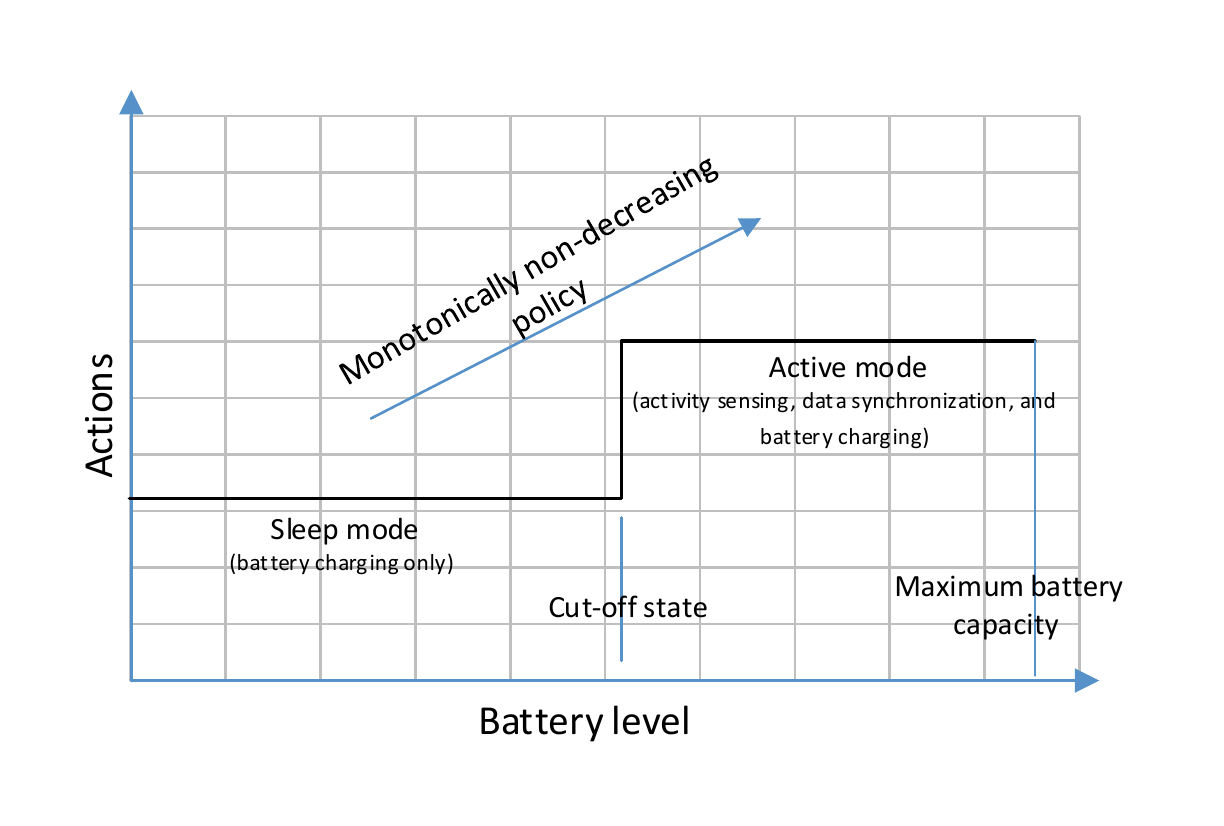}
\par\end{centering}

\caption{Threshold activity tracking policy with two actions corresponding to the sleep and active modes. The sleep mode is preferred at low battery levels while data sensing and synchronization are performed at high levels. \label{fig:threshold_policy_example}}

\end{figure}

Our optimization problem is intentionally designed to derive a threshold activity tracking policy. A \emph{threshold activity tracking policy}, which is an MDP solution that follows a monotone pattern with system states, facilitates problem solution and implementation~\cite{puterman2005markov}.  In this paper, the optimal policy of the  MDP is shown to be a threshold in the battery level. Figure~\ref{fig:threshold_policy_example} shows the desired structure of the activity tracking policy. A \emph{cut-off state} of a threshold policy is the battery level beyond which the selected action is increased to a new value. This is important due to the following benefits:
\begin{itemize}
\item \emph{Fast adaptation of online policy learning}: Solution methods for computing an optimal activity tracking policy can be customized to explore the threshold structure of the intended policy. This is important as activity statistics are unknown a priori, varying with time, and inconsistent for different users. Specifically, this customization significantly improves the convergence of conventional solution methods~\cite{djonin2007learning,kunnumkal2008exploiting}. 
\item \emph{Low memory and communication overheads}: Saving the threshold policy into the mobile device's memory in a compact form is significantly efficient, as threshold cut-off states are sufficient for policy execution. This requires low memory footprint compared with unstructured policies which are saved using look-up tables with state-action pairs. Similarly, the overhead in transferring the learned policy between the system components is also minimized by only sending the threshold cut-off states.
\item \emph{Simple implementation}: The threshold policy helps in developing simple and lightweight algorithms. Clearly, the selection of optimal online actions can be done by comparing the system state with the cut-off value, e.g., using a simple if-then-else statement, and no look-up search is needed.
\end{itemize}


\section{Unconstrained Activity Tracking Policy: A Lagrange Relaxation Approach}\label{sec:unconstrained_MDP}

In this section, the CMDP formulation is transformed into its unconstrained MDP form using a Lagrange multiplier. Firstly, the unconstrained MDP is required to prove the threshold structure of the CMDP policy. Particularly, utilizing the threshold structure of the CMDP for the mobile sensing in (\ref{eq:cmdp_1})-(\ref{eq:cmdp_4}) is complex due to the data usage term. Secondly, the complexity of the value iteration algorithm is lower than that of the algorithm to solve the LP problem~\cite{littman1995complexity}. Therefore, the CMDP formulation must be first transformed into an unconstrained MDP. We adopt the transformation approach that relies on using the Lagrange multiplier algorithm~\cite{beutler1985optimal}. The Lagrangian relaxation approach introduces a Lagrange multiplier $\lambda$ and the resulting \emph{Lagrangian error function} is defined as follows:
\begin{equation}
c(\psi,\delta;\lambda)=c(\psi,\delta)+\lambda d(\psi,\delta),\label{eq:lagrangian_cost_function}
\end{equation}
where $\lambda>0$. Accordingly, the \emph{Lagrangian average error} $\mathcal{J}\left(\pi;\lambda\right)$ is given by
\begin{equation}
\mathcal{J}\left(\pi;\lambda\right)=\lim_{N\rightarrow\infty}\textrm{sup}\frac{1}{N}\sum_{n=1}^{N}\mathbb{E}(c(\psi^{n},\delta^{n})+\lambda d(\psi^{n},\delta^{n})).\label{eq:Lagrangian_cost}
\end{equation}
This Lagrangian average error enables the solution of the CDMP problem using any of the conventional MDP solution methods such as the value iteration or policy iteration algorithms. Thus, the MDP activity tracking policy is found by minimizing (\ref{eq:Lagrangian_cost}) such that
\begin{equation}
\pi_{\textrm{MDP}}^{*}(\psi)=\arg\inf \mathcal{J}\left(\pi;\lambda\right).\label{eq:unconstrained_mdp_solution}
\end{equation}
In the following, we discuss solution methods of the unconstrained MDP in (\ref{eq:unconstrained_mdp_solution}). In particular, the unconstrained problem is formulated as an LP using which the optimal Lagrange multiplier $\lambda^{*}$ is selected. Then, given $\lambda^{*}$, the problem is solved using the value iteration algorithm.

\subsection{Solution Using Linear Programming}

It is important to note that $\pi_{\textrm{MDP}}^{*}$ can still be solved using LP solvers as in (\ref{eq:cmdp_1})-(\ref{eq:cmdp_4}) after dropping (\ref{eq:cmdp_2}) and replacing $c(\psi,\delta)$ in (\ref{eq:cmdp_1}) with $c(\psi,\delta;\lambda)$. The resulting LP problem is as follows:
\begin{eqnarray}
\min_{\phi(\psi,\delta)} && \sum\limits _{\psi\in\Psi}\sum\limits _{\delta\in\Delta}\phi(\psi,\delta)c(\psi,\delta;\lambda),\label{eq:cmdp_1_1}\\
\textrm{s.t.} & & \sum\limits _{\delta\in\Delta}\phi(\psi',\delta)=\sum_{\psi\in\Psi}\limits\sum\limits _{\delta\in\Delta}\phi(\psi,\delta)\mathbb{P}(\psi'|\psi,\delta),\label{eq:cmdp_3_1}\\
 & & \sum\limits _{\psi\in\Psi}\sum\limits _{\delta\in\Delta}\phi(\psi,\delta)=1,\phi(\psi,\delta)\geq0,\label{eq:cmdp_4_1}
\end{eqnarray}
where $\psi'\in\Psi$. A special attention should be given to the selection of the Lagrange multiplier $\lambda$ to ensure that the resulting unconstrained activity tracking policy is an accurate transformation of the optimal CMDP activity tracking policy. Therefore, finding an optimal Lagrange value $\lambda^{*}$ is discussed next.

\subsection{Finding $\lambda^{*}$}

Clearly, there is a strong connection between the chosen value of the Lagrange multiplier $\lambda$ and the cellular data limit $D$. Explicitly, $\lambda^{*}$ is found as follows~\cite{beutler1985optimal,sennott1993constrained}:
\begin{equation}
\lambda^{*}=\inf\left\{ \lambda:\mathcal{D}\left(\pi^{*};\lambda\right)\leq D\right\} .\label{eq:lagrange_optimality}
\end{equation}
An optimal value of $\lambda^{*}$ can be obtained by iterative methods as shown in Algorithm~\ref{alg:lagrange_learning} such that the policy always meets the cellular data limit $D$. At each iteration $i$ of Algorithm~\ref{alg:lagrange_learning}, Step~\ref{alg:lagrange_learning_solve} solves the LP problem given in~(\ref{eq:cmdp_1_1})-(\ref{eq:cmdp_4_1}) based on the Lagrange multiplier estimation $\lambda_{i}$. This solution is used to find the corresponding data usage $\mathcal{D}\left(\pi^{*};\lambda\right)$ as in Step~\ref{alg:lagrange_learning_constraint} which updates the Lagrange multiplier in Step~\ref{alg:lagrange_learning_update}. This update rule in Step~\ref{alg:lagrange_learning_update} is based on the Robbins\textendash{}Monro algorithm for stochastic approximation which has been used in previous studies for the Lagrange multiplier estimation~\cite{djonin2007learning,ngo2010monotonicity}. The algorithm terminates when the difference between two estimations of the Lagrange multiplier is below a small error value $\epsilon$.

\begin{algorithm}
\begin{algorithmic}[1] 
\Require{$\Psi,\Delta,\mathbb{P},c(\cdot),d(\cdot)$}
\Ensure{$\lambda^{*}$}
\State{Initialize an iteration counter $i=0$}
\State{Initialize $\lambda_{0}$ to a random number greater than $0$}
\State{Solve (\ref{eq:cmdp_1_1})-(\ref{eq:cmdp_4_1}) using LP to find $\phi_{\lambda_{i}}^{*}(\psi,\delta)$, $\psi\in\Psi$ and $\delta\in\Delta$\label{alg:lagrange_learning_solve}}
\State{Find current data usage $\mathcal{D}\left(\pi^{*};\lambda_{i}\right)=\sum\limits _{\psi\in\Psi}\sum\limits _{\delta\in\Delta}\phi_{\lambda_{i}}^{*}(\psi,\delta)d(\psi,\delta)$\label{alg:lagrange_learning_constraint}}
\State{Update $\lambda_{i+1}=\lambda_{i}+\frac{1}{\sqrt{i+1}}\left(\mathcal{D}\left(\pi^{*};\lambda_{i}\right)-D\right)$\label{alg:lagrange_learning_update}}
\State{if $\left|\lambda_{i+1}-\lambda_{i}\right|<\varepsilon$, terminate and go to Step \ref{alg:lagrange_learning_exit}}
\State{Increment counter $i=i+1$ and go to Step \ref{alg:lagrange_learning_solve}}
\State{\Return{$\lambda^{*}=\min\limits_{0\leq j\leq i+1}\left\{ \lambda_{j}:\mathcal{D}\left(\pi^{*};\lambda_{j}\right)\leq D\right\}$}\label{alg:lagrange_learning_exit}}
\end{algorithmic}

\caption{Lagrange multiplier estimation of the unconstrained activity tracking policy.\label{alg:lagrange_learning}}
\end{algorithm}

\subsection{Discounted Cost Solutions}

$\pi_{\textrm{MDP}}^{*}$ can also be found using discounted cost, infinite horizon MDP solution methods~\cite{puterman2005markov}. Based on the previous unconstrained formulation, the \emph{expected total discounted error} for a policy $\pi$ and a discount factor $\beta$, where $0\leq\beta<1$, can be expressed as follows: 
\begin{equation}
J\left(\pi;\lambda,\beta\right)=\lim_{N\rightarrow\infty}\textrm{sup}\left(\mathbb{E}\left[\sum\limits _{n=1}^{N}\beta^{n-1}c(\psi,\delta;\lambda)\right]\right),
\end{equation}
which can be solved using the value iteration algorithm. Accordingly, the optimal value function $v\left(\psi;\lambda,\beta\right)$ for each state $\psi\in\Psi$ is
\begin{multline}
v\left(\psi;\lambda,\beta\right)=\\
\min_{\delta\in\triangle}\left\{ c(\psi,\delta;\lambda)+\beta\sum\limits _{\psi'\in\Psi}\mathbb{P}(\psi'|\psi,\delta)v\left(\psi';\lambda,\beta\right)\right\},\label{eq:bellman_equation}
\end{multline}
which is the solution of the Bellman equation with stationary policy $\pi_{\text{MDP}}^{*}$ defined as follows:
\begin{multline}
\pi_{\text{MDP}}^{*}=\\
\arg\min_{\delta\in\triangle}\left\{ c(\psi,\delta;\lambda)+\beta\sum\limits _{\psi'\in\Psi}\mathbb{P}(\psi'|\psi,\delta)v\left(\psi';\lambda,\beta\right)\right\} .\label{eq:bellman_equation_soln}
\end{multline}
The Bellman equation can be recursively solved using the value iteration algorithm. In particular, the \emph{value function} $v(\psi;\lambda,\beta)$ and the \emph{state-action cost function} (or called the Q-function) $Q(\psi,\delta;\lambda,\beta)$ are arbitrarily initialized and then updated in each iteration of the value iteration algorithm as follows:
\begin{multline}
v^{i+1}(\psi;\lambda,\beta)=\\
\min_{\delta\in\triangle}\left\{ c(\psi,\delta;\lambda)+\beta\sum\limits _{\psi'\in\Psi}\mathbb{P}(\psi'|\psi,\delta)v^{i}\left(\psi';\lambda,\beta\right)\right\},\label{eq:value_function}
\end{multline}
\begin{multline}
Q^{i+1}(\psi,\delta;\lambda,\beta)=\\
c(\psi,\delta;\lambda)+\beta\sum\limits _{\psi'\in\Psi}\mathbb{P}(\psi'|\psi,\delta)v^{i}\left(\psi';\lambda,\beta\right),\label{eq:state_action_cost}
\end{multline}
where $i\in\left\{ 0,1,2,\ldots\right\} $ is an iteration counter of the algorithm, $v^{i}(\psi;\lambda,\beta)$ is the minimum achievable cost value for state $\psi$ at iteration $i$, and $Q^{i}(\psi,\delta;\lambda,\beta)$ is the minimum cost value for taking action $\delta$ during state $\psi$ at time $i$. Here, it is important to note that the value function is obtained for each state, while the state-action cost function is found for each state-action pair. After convergence, the policy $\pi_{\text{MDP}}^{*}$ is deterministic and stationary~\cite{puterman2005markov,bertsekas2012dynamic}. Conversely, $\pi_{\textrm{CMDP}}^{*}$ is a randomized, stationary policy~\cite{altman1999constrained}. Thereby, a key note is that $\pi_{\textrm{CMDP}}^{*}$ with the average cost is not directly estimated by $\pi_{\text{MDP}}^{*}$ that uses the discounted cost. Instead, $\pi_{\textrm{CMDP}}^{*}$ will be shown in the next section to be a randomized mixture of two perturbed policies of $\pi_{\text{MDP}}^{*}$.

To this end, the constrained MDP activity tracking policy was transformed as a discounted cost and unconstrained MDP by introducing a Lagrange error function given in (\ref{eq:lagrangian_cost_function}). The unconstrained transformation was solved using conventional method, such as the value iteration algorithm, which meets the Bellman equation (\ref{eq:bellman_equation}). In the following section, the threshold structure of the activity tracking policy is analytically proved by using the discounted cost MDP formulation and the concept of submodularity.

\section{Randomized Threshold Activity Tracking Policy: Monotonicity Analysis}\label{sec:structure_analysis}

This section proves the monotone, threshold structure of the proposed activity tracking policy. In summary, this section includes the following contributions:
\begin{itemize}
\item The concept of submodularity~\cite{topkis1998supermodularity} is used to prove the monotone structure of the unconstrained Lagrange formulation of the activity tracking policy. This analysis is based on the discounted cost MDPs in (\ref{eq:bellman_equation}). This proof requires two major steps: (a)~the value function $v(\psi;\lambda,\beta)$ must be monotone, and (b)~the state-action cost function $Q(\psi,\delta;\lambda,\beta)$ must be submodular.
\item After proving the threshold structure of the unconstrained MDPs, the CMDP policy is shown to be a mixture of two threshold MDP policies. Accordingly, this proves the threshold structure for the optimal activity tracking policy.
\item A threshold Q-learning method that exploits the threshold structure of the activity tracking policy is presented. This method enables a fast convergence compared with the conventional Q-learning method.
\end{itemize}

\subsection{Threshold Structure of the Unconstrained Policy}

To prove the threshold structure, this section follows the same approach as in~\cite{djonin2007mimo,ngo2010monotonicity}. Firstly, the concept of submodularity will be defined to prove the monotone structure of a policy as submodularity is a sufficient condition for proving threshold policy structure~\cite{puterman2005markov}. In summary, our main objective is to prove that the state-action cost function $Q(\psi=\left[u,e,b\right],\delta;\lambda,\beta)$ is submodular in $(b\in\mathcal{B},\delta\in\triangle)$, and hence the optimal activity tracking policy is monotonically non-decreasing in $b$.
\begin{defn}
For any two sets $\mathcal{X}\subseteq\mathbb{R}$ and $\mathcal{Y\subseteq\mathbb{R}}$, a function $f(\cdot)$ that is defined as $f:\mathcal{X}\times\mathcal{Y}\rightarrow\mathbb{R}$ is called \emph{submodular} in $(x\in\mathcal{X},y\in\mathcal{Y})$ if the inequality condition $f(x_{1},y_{1})+f(x_{2},y_{2})\leq f(x_{1},y_{2})+f(x_{2},y_{1})$ holds for all $x_{1}\geq x_{2}$, $y_{1}\geq y_{2}$, $x_{1},x_{2}\in\mathcal{X}$, and $y_{1},y_{2}\in\mathcal{Y}$.
\end{defn}
This definition is important as the submodularity of $f(\cdot)$ is a sufficient condition for the non-decreasing monotonicity of $y=\arg\min f(x,y)$ \cite{topkis1998supermodularity}.

\begin{lem}\label{lem:value_monotonicity}
For a given optimal Lagrange multiplier $\lambda^{*}>0$ and a discount factor $\beta\in[0,1)$, the optimal value function $v(\psi;\lambda,\beta)$ of the mobile sensing is monotonically non-decreasing in the battery level $b$.
\end{lem}
\begin{IEEEproof}
See the Appendix.
\end{IEEEproof}

\begin{thm}\label{thm:monotone_policy}
For a given optimal Lagrange multiplier $\lambda^{*}>0$ and a discount factor $\beta\in[0,1)$, the optimal discounted cost MDP policy $\pi^{*}_{\text{MDP}}$ of the mobile sensing is monotone and does not decrease as the battery level $b$ increases.\end{thm}
\begin{IEEEproof}
See the Appendix.
\end{IEEEproof}

Based on Theorem \ref{thm:monotone_policy} and for a given threshold value $b_{\text{cut}}\in\mathcal{B}$, the optimal discount policy $\pi_{\text{MDP}}^{*}$ can be written in the \emph{compact form} as follows:
\begin{equation}
\pi_{\text{MDP}}^{*}(\psi=\left[u,e,b\right])=\begin{cases}
0, & 0< b\leq b_{\text{cut}},\\
1, & b_{\text{cut}}\leq b\leq B.
\end{cases}\label{eq:threshold_policy}
\end{equation}
This compact form of the threshold policy facilitates the development of the activity tracking policy as discussed in Section~\ref{sec:mobile_sensing}. $\pi_{\text{MDP}}^{*}(\psi)$ is called a \emph{binary threshold policy} as it only selects between two possible actions $\delta_{0}$ and $\delta_{1}$. The intuition of the non-decreasing structure is that when the battery level is higher, the mobile device will take the activation action due to the lower cost.

\subsection{Threshold Structure of the Constrained Policy}

The threshold structure of the discounted cost MDP problem is proved as in Theorem~\ref{thm:monotone_policy}. Correspondingly, the infinite horizon average cost CMDP problem with a single constraint has an optimal policy $\pi_{\textrm{CMDP}}^{*}$ that is a mixture of two threshold MDP policies in the form~\cite{beutler1985optimal,sennott1993constrained} presented as follows:
\begin{equation}
\pi_{\textrm{CMDP}}^{*}=\gamma\pi_{\textrm{MDP}}^{+}+(1-\gamma)\pi_{\textrm{MDP}}^{-},\label{eq:mixed_policy}
\end{equation}
where $\pi_{\textrm{MDP}}^{+}$ and $\pi_{\textrm{MDP}}^{-}$ are the stationary discounted cost MDP policies for perturbed values of $\lambda^{+}=\lambda^{*}+\Delta\lambda$ and $\lambda^{-}=\lambda^{*}-\Delta\lambda$, where $\Delta\lambda$ is the perturbation value of $\lambda$. $\gamma\in[0,1]$ is the probability of selecting $\pi_{\textrm{MDP}}^{+}$ and $1-\gamma$ is for selecting $\pi_{\textrm{MDP}}^{-}$. Even though the MDP policy is deterministic and stationary, (\ref{eq:mixed_policy}) is important as it enables the randomized selection of optimal actions in a randomized manner, i.e., (\ref{eq:mixed_policy}) stochastically selects actions as a CMDP policy. Here, the probability $\gamma$ can be calculated using the rule $\gamma=\frac{\mathcal{D}\left(\pi_{\textrm{MDP}}^{-}\right)-D}{\mathcal{D}\left(\pi_{\textrm{MDP}}^{-}\right)-\mathcal{D}\left(\pi_{\textrm{MDP}}^{+}\right)}.$

\subsection{Fast Adaptation of Q-Learning}\label{sub:structured_q_learning}

In online activity tracking, the transition probabilities between user activities could be (i)~unknown at design time, (ii)~changing over time for the same user, or (iii)~distinct for different users. Therefore, an online algorithm that can adapt with the changing parameters is required.


The Q-learning algorithm~\cite{watkins1992q} is widely considered as one of the most powerful \emph{model-free} methods of reinforcement learning. The gradual learning feature of the Q-learning allows the policy formulation of an activity-aware mobile device when the transition matrix $\mathbb{P}\left(\psi'|\psi,\delta\right)$ is unknown. Additionally, this enables the customization of the activity tracking policy to match each user's temporal behavior. Q-learning estimates (\ref{eq:state_action_cost}) using stochastic approximation by starting with a random (or zero) approximations of each state-action cost $\left\{ Q^{0}(\psi,\delta;\lambda,\beta):\forall\psi\in\Psi\text{ and }\delta\in\Delta\right\} $. Then, the conventional Q-learning update rule can be expressed as follows:
\begin{multline}
Q^{i+1}(\psi,\delta;\lambda,\beta)=Q^{i}(\psi,\delta;\lambda,\beta)+\frac{1}{\sqrt{i+1}}\\
\times\left[c(\psi,\delta;\lambda)+\beta\min_{\delta'\in\Delta}Q^{i}(\psi',\delta';\lambda,\beta)-Q^{i}(\psi,\delta;\lambda,\beta)\right],\label{eq:q-learning}
\end{multline}
where $\beta\in[0,1)$ is a discount factor, and $i\in\left\{ 1,2,\ldots,L\right\} $ is an iteration counter that is limited to an upper bound $L$, e.g., $L=10^{5}$ iterations. (\ref{eq:q-learning}) is a greedy update rule that the lowest-cost action in the next state is used to update the state-action cost factor $Q^{i+1}(\psi,\delta;\lambda,\beta)$ of the current state.

A known limitation of the conventional Q-learning algorithm is the long sequence of iterations required in practical applications~\cite{djonin2007learning,kunnumkal2008exploiting}. Recall that the monotonically non-decreasing structure of the activity tracking policy was proven in Theorem~\ref{thm:monotone_policy}. This structure enables a faster convergence of the Q-learning algorithm by (a)~initializing the state-action cost factor of all states such that $Q^{0}(\psi=[u,e,b],\delta;\lambda,\beta)<Q^{0}(\psi=[u,e,b+1],\delta;\lambda,\beta)$ for all states $\psi\in\Psi$ and action $\delta\in\Delta$ (see~\cite{kunnumkal2008exploiting}), and (b)~projecting the final policy such that it preserves the threshold structure. These steps limit the search of the optimal policy to a subset of all possible solutions, and hence avoid the brute-force search in conventional Q-learning.


\section{Numerical Results and Discussion}\label{sec:numerical_validation}

This section presents numerical analysis of the optimal activity tracking policy. Firstly, parameter settings using a real-world dataset are presented. Then, the threshold structure validation results are summarized. Finally, performance measures and performance evaluation of the CMDP policy is given.

\subsection{Parameter Setting}

\begin{figure}
\begin{centering}
\includegraphics[width=1.0\columnwidth]{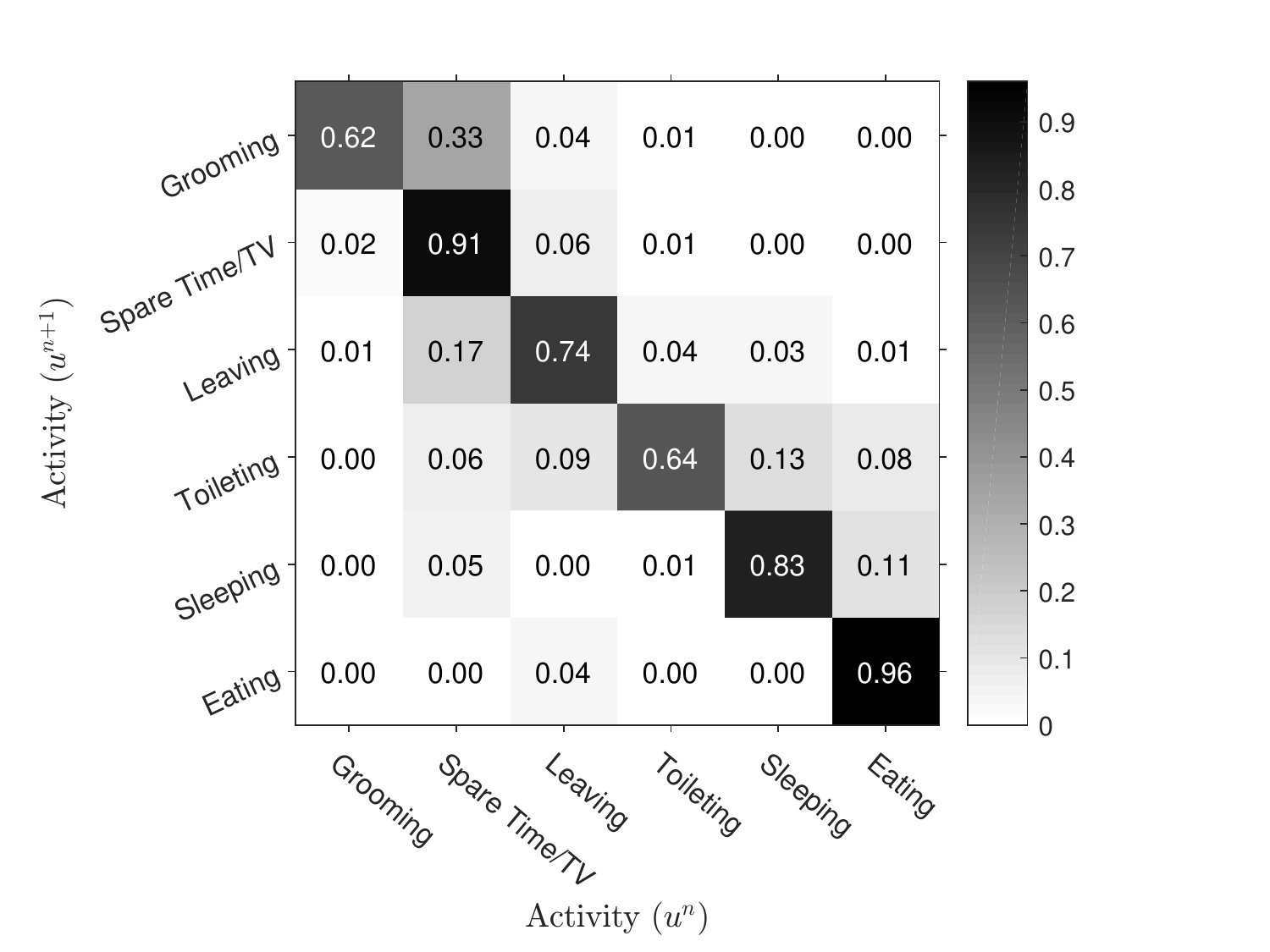}
\par\end{centering}

\caption{Transition probabilities $\mathbb{P}_{\textrm{user}}$ between user activities.\label{fig:transition_probability}}
\end{figure}

We run experiments while using a real-world dataset~\cite{ordonez2013activity} to extract the system parameters. Specifically, we consider an activity tracking scenario of six activities in daily home routines such that $\mathcal{U}$ is defined as follows:
\begin{multline}
\mathcal{U}= \big\{0:\text{grooming},1:\text{spare time/watching TV},\\
2:\text{leaving},3:\text{sleeping},4:\text{toileting/showering}, 5:\text{eating}\big\}.
\end{multline}
Unless otherwise stated, the battery charging can be in one of two modes: (i)~a charging mode when the mobile device is connected to a wired or wireless charger with a probability of $\mathbb{P}(e=1)=0.15$, and (ii)~no-charging mode with a probability of $\mathbb{P}(e=0)=\left(1-\mathbb{P}(e=1)\right)=0.85$. For convenience, we denote the user transition probability matrix as $\mathbb{P}_{\textrm{user}}=\left[P(u'|u),\forall u\in\mathcal{U}\text{ and }u'\in\mathcal{U}\right]$ which is given in Figure~{\ref{fig:transition_probability}}. The immediate detection error in~(\ref{eq:cost_function}) is defined such that $c(\psi=[u=0,e,b],\delta_{1})=0.28$, $c(\psi=[u=1,e,b],\delta_{1})=0.25$, $c(\psi=[u=2,e,b],\delta_{1})=0.18$, $c(\psi=[u=3,e,b],\delta_{1})=0.12$, $c(\psi=[u=4,e,b],\delta_{1})=0.1$, and $c(\psi=[u=5,e,b],\delta_{1})=0.08$. The user activities are not tracked during the sleep mode such that $c(\psi,\delta_{0})=1$. The data usage function is defined as follows:
\[
d(\psi=[u,e,b],\delta)=\begin{cases}
1, & \delta=1\text{ and }b>0,\\
0, & \text{otherwise}.
\end{cases}
\]
This indicates that one data packet is generated during the active mode. The connectivity probabilities in (\ref{eq:coverage_probability}) are $g(\psi=[u=0,e,b],\delta)=0.5\delta$, $g(\psi=[u=1,e,b],\delta)=0.55\delta$, $g(\psi=[u=2,e,b],\delta)=0.6\delta$, $g(\psi=[u=3,e,b],\delta)=0.65\delta$, $g(\psi=[u=4,e,b],\delta)=0.68\delta$, and $g(\psi=[u=5,e,b],\delta)=0.7\delta$. Clearly, data transmission cannot be performed during the sleep mode $\text{g}(\psi=[u,e,b],\delta_{0})=0$. Finally, the discount factor in the discounted cost MDP formulation is $\beta=0.99$ to ensure high precision solutions.

\subsection{Threshold Structure}

\begin{figure}
\begin{centering}
\subfloat[\label{fig:policy_cmdp_lp}]{\begin{centering}
\includegraphics[width=0.9\columnwidth,trim=0cm 1cm 0cm 0cm]{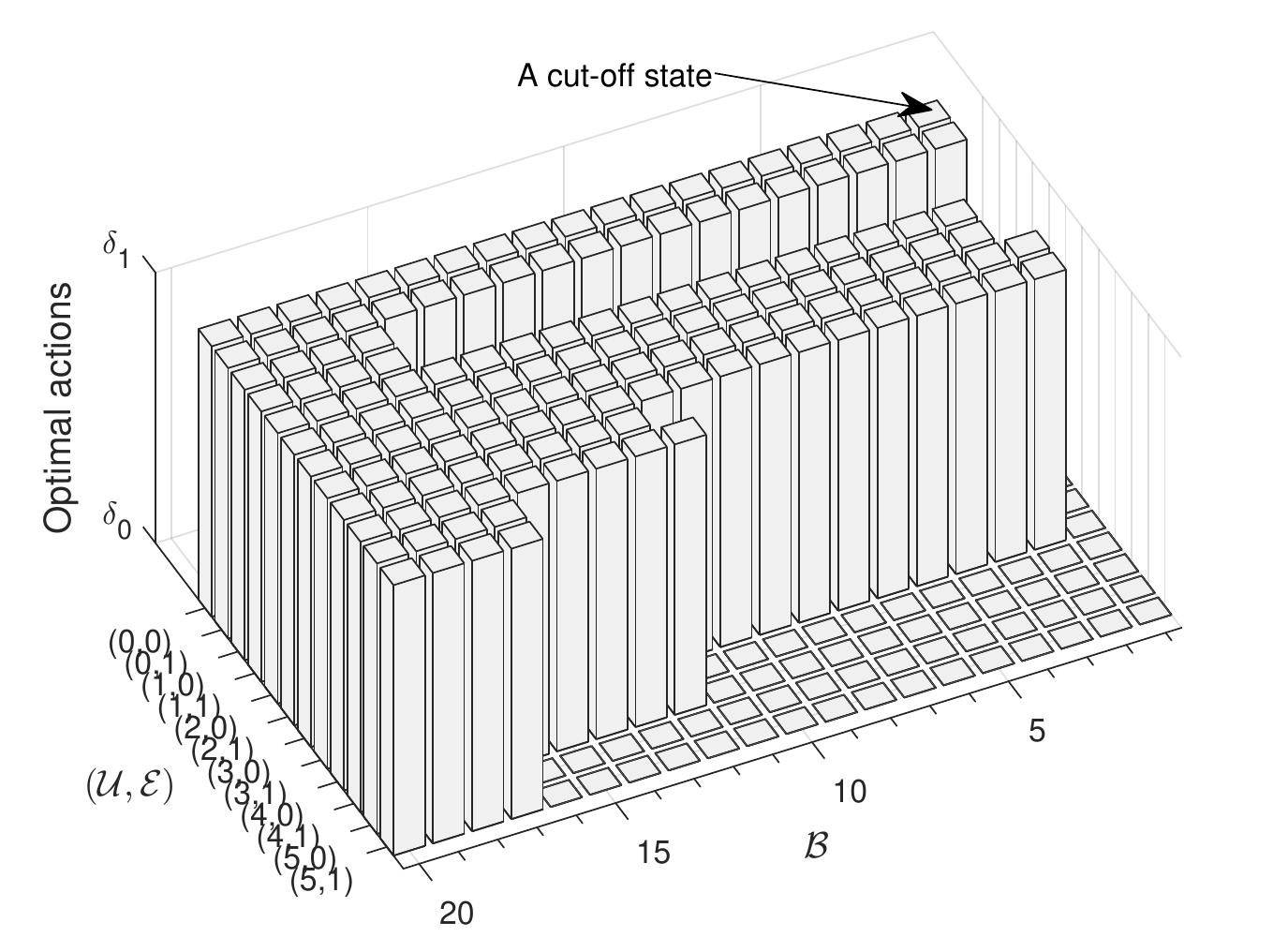}
\par\end{centering}}\\
\subfloat[\label{fig:fig:policy_mdp_lp}]{\begin{centering}
\includegraphics[width=0.9\columnwidth,trim=0cm 1cm 0cm 0.4cm]{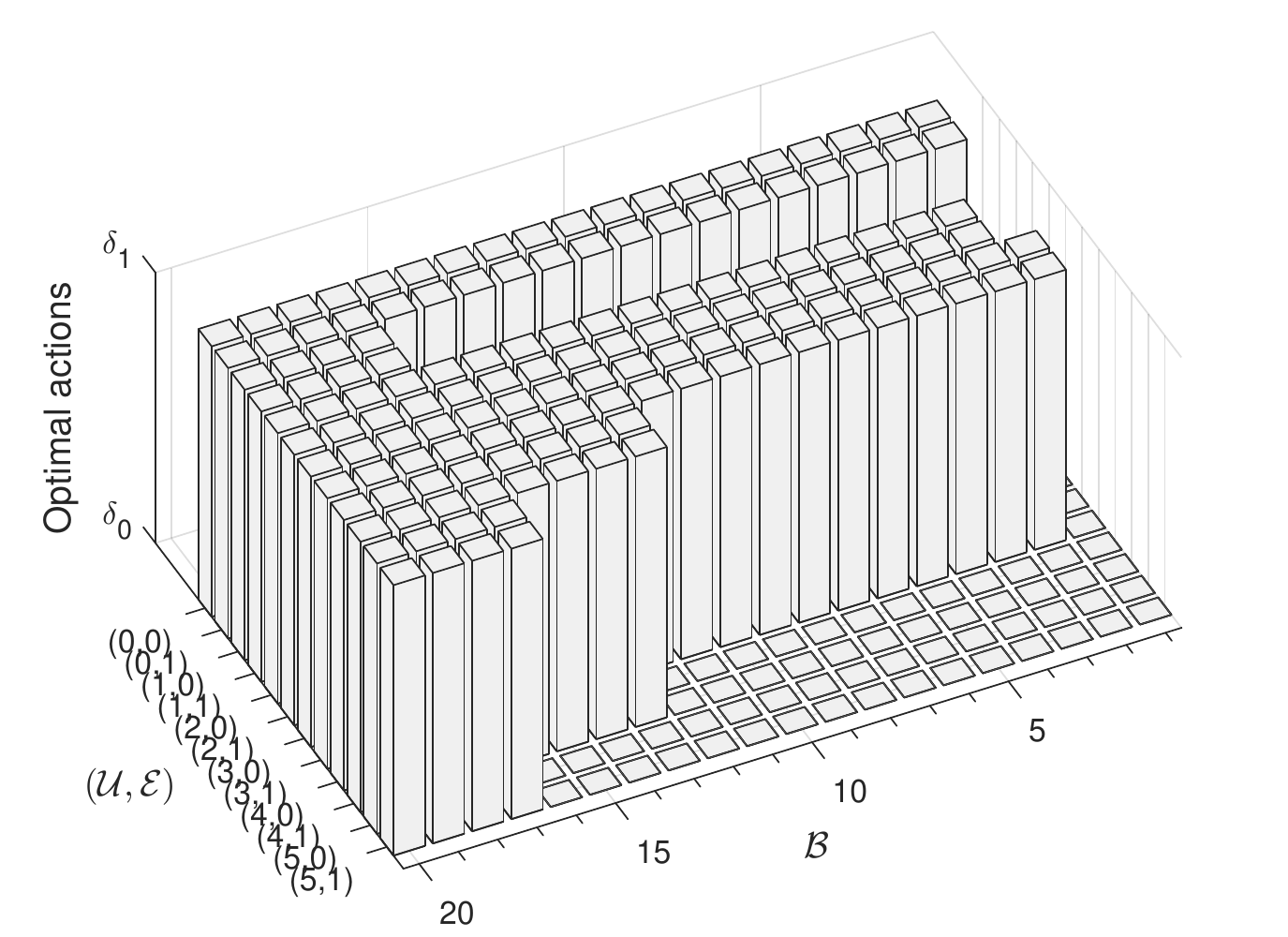}
\par\end{centering}}\\
\subfloat[\label{fig:fig:policy_mdp_vi}]{\begin{centering}
\includegraphics[width=0.9\columnwidth,trim=0cm 0.8cm 0cm 0.4cm]{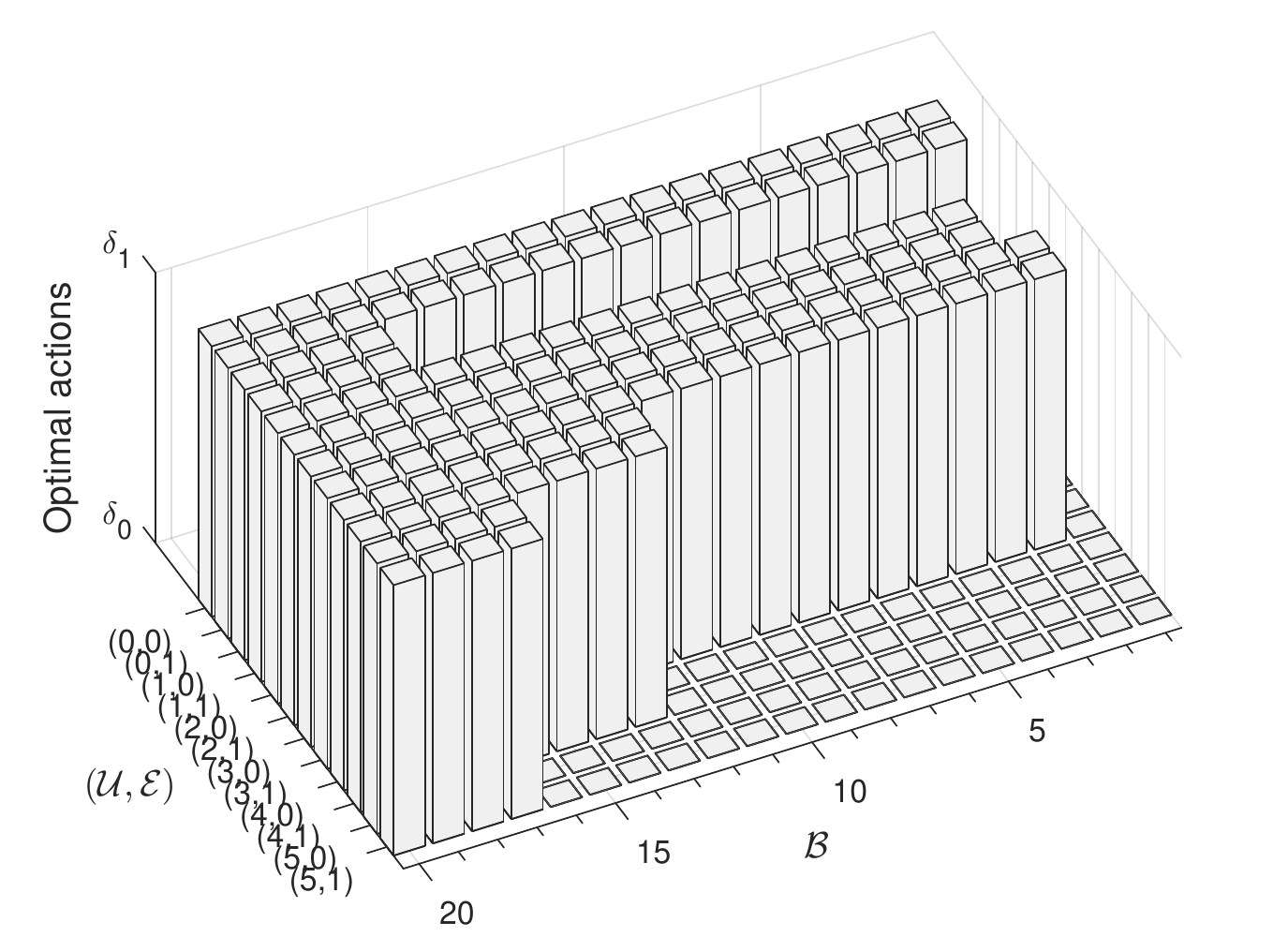}
\par\end{centering}}
\par\end{centering}

\caption{\label{fig:threshold_policy}Threshold policies for an activity-aware mobile device. Recall that $\mathcal{U},\mathcal{E}$, and $\mathcal{B}$ are the user activity, energy charging, and battery level state spaces of the mobile device, respectively. (a)~The constrained MDP solution to the mobile sensing problem using LP, (b)~the unconstrained MDP solution using LP with $\lambda=0.128$, and (c)~the unconstrained MDP  solution using value iteration with $\lambda=0.128$.}
\end{figure}

We first analyze the threshold structure of the activity tracking policy. The feasible battery levels are $\mathcal{B}=\left\{ 0,\ldots,20\right\} $, and the cellular data limit is fixed as $D=0.25$. This indicates that the system senses the user activities in one fourth of the total decision epochs $\mathcal{N}$. We use the following steps to obtain the optimal CMDP policy and its unconstrained MDP estimation: (i)~the problem is solved using the CMDP formulation in (\ref{eq:cmdp_1})-(\ref{eq:cmdp_4}) and LP, (ii)~the CMDP problem is then transformed into the unconstrained MDP form given in (\ref{eq:cmdp_1_1})-(\ref{eq:cmdp_4_1}), (iii)~the optimal Lagrange multiplier value $\lambda^{*}=0.128$ is found using Algorithm~\ref{alg:lagrange_learning}, and (iv)~the unconstrained problem is solved using the value and policy iteration algorithms. Figure~\ref{fig:threshold_policy} shows the resulting policies of the constrained and unconstrained MDP formulations. Two important results from Figure~\ref{fig:threshold_policy} can be highlighted as follows:
\begin{enumerate}
\item The optimal activity tracking policy has a threshold structure. In particular, the policy is a threshold policy and is monotonically non-decreasing in the battery level $b$. This observation represents the outcome from Theorem~\ref{thm:monotone_policy}. In this case, the optimal actions taken by the mobile device change from $\delta_{0}$ to $\delta_{1}$ as the battery level increases. The cut-off states are the only data required by the mobile device for selecting the optimal actions.
\item The discounted cost MDP solution using (\ref{eq:mixed_policy}) provides an accurate transformation of the CMDP problem. This simulation result is consistent with the theoretical analysis in Sections~\ref{sec:mobile_sensing} and \ref{sec:structure_analysis}, where the CMDP policy is a randomized mixture of the discounted cost MDP policies.
\end{enumerate}

\subsection{The Impact of Setting the Lagrange Multiplier}

\begin{figure}
\begin{centering}
\includegraphics[width=1.0\columnwidth,trim=0cm 0cm 0cm 0.0cm]{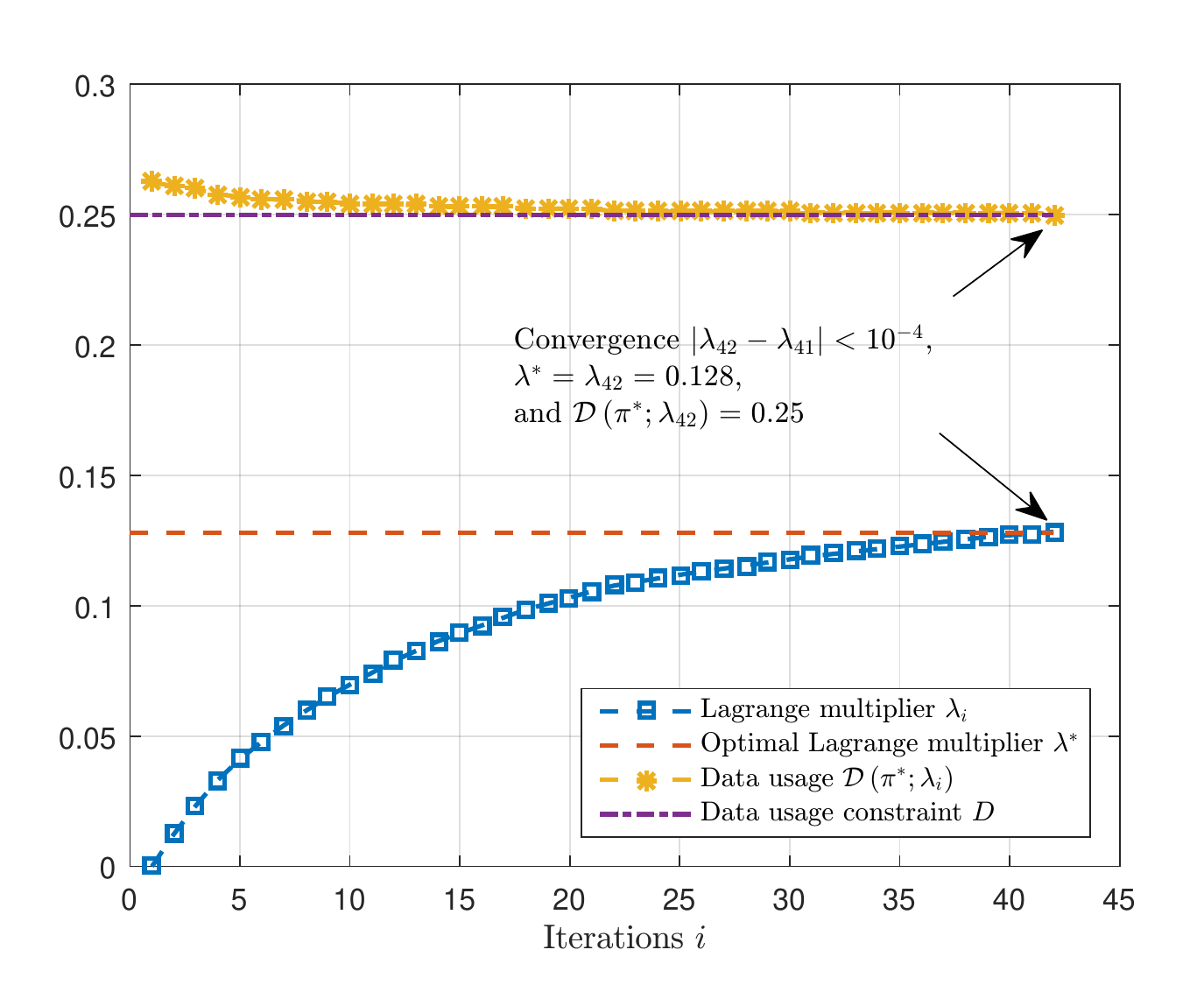}
\par\end{centering}

\caption{Setting the optimal Lagrange multiplier $\lambda^{*}$ using Algorithm~\ref{alg:lagrange_learning}.\label{fig:lagrange_update}}
\end{figure}

Figure~\ref{fig:lagrange_update} shows the Lagrange multiplier $\lambda$ updates over iterations of Algorithm~\ref{alg:lagrange_learning} with $\epsilon=10^{-4}$ and $D=0.25$. Based on this experiment, the optimal value is found as $\lambda^{*}=0.128$ which satisfies the optimality condition in (\ref{eq:lagrange_optimality}). The data usage is found at each iteration as $\mathcal{D}\left(\pi^{*};\lambda_{i}\right)=\sum\limits _{\psi\in\Psi}\sum\limits _{\delta\in\Delta}\phi_{\lambda_{i}}^{*}(\psi,\delta)d(\psi,\delta)$.

\begin{figure}
\begin{centering}
\includegraphics[width=0.9\columnwidth,trim=0cm 0cm 0cm 0.5cm]{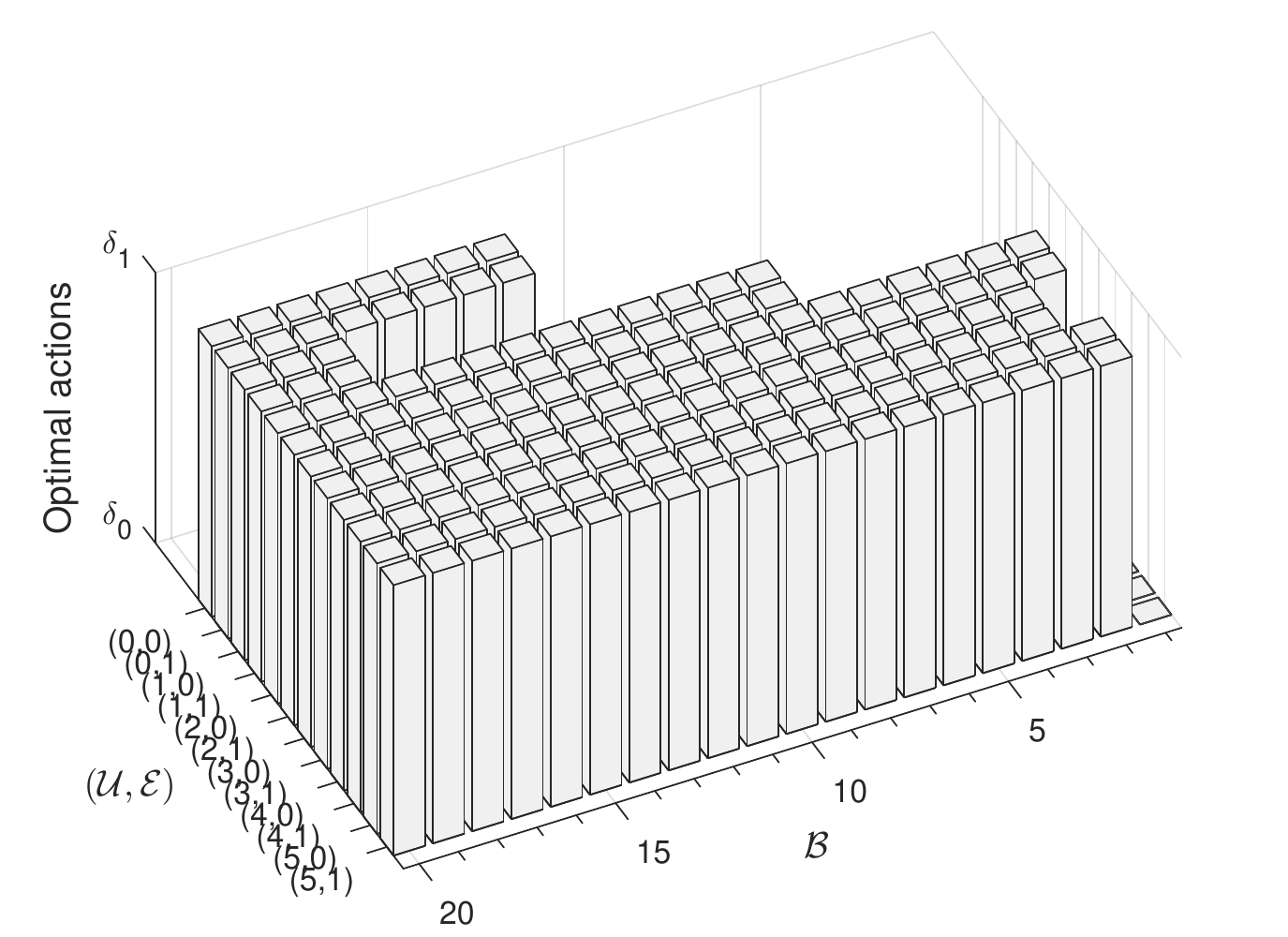}
\par\end{centering}
\caption{The unconstrained MDP solution using value iteration with an incorrect Lagrange multiplier of $\lambda=0.25$ resulting in a poor estimation solution.\label{fig:fig:policy_mdp_vi_wrong}}
\end{figure}

The Lagrange optimal value selection is critical for the accuracy of the unconstrained discounted cost MDP solutions. Figure~\ref{fig:fig:policy_mdp_vi_wrong} shows the policy when the Lagrangian multiplier is incorrectly set as $\lambda=0.25$. Intuitively, an incorrect value of the Lagrange multiplier results in a poor estimation of the CMDP optimal policy. This because the Lagrangian error function defined in (\ref{eq:lagrangian_cost_function}) does not accurately capture the data usage constraint of the CMDP activity tracking policy given in (\ref{eq:cmdp_1})-(\ref{eq:cmdp_4}).

\subsection{Fast Adaptation of Q-learning}

\begin{figure*}
\begin{centering}
\subfloat[\label{fig:structured_q_learning_1}]{\begin{centering}
\includegraphics[width=0.9\columnwidth]{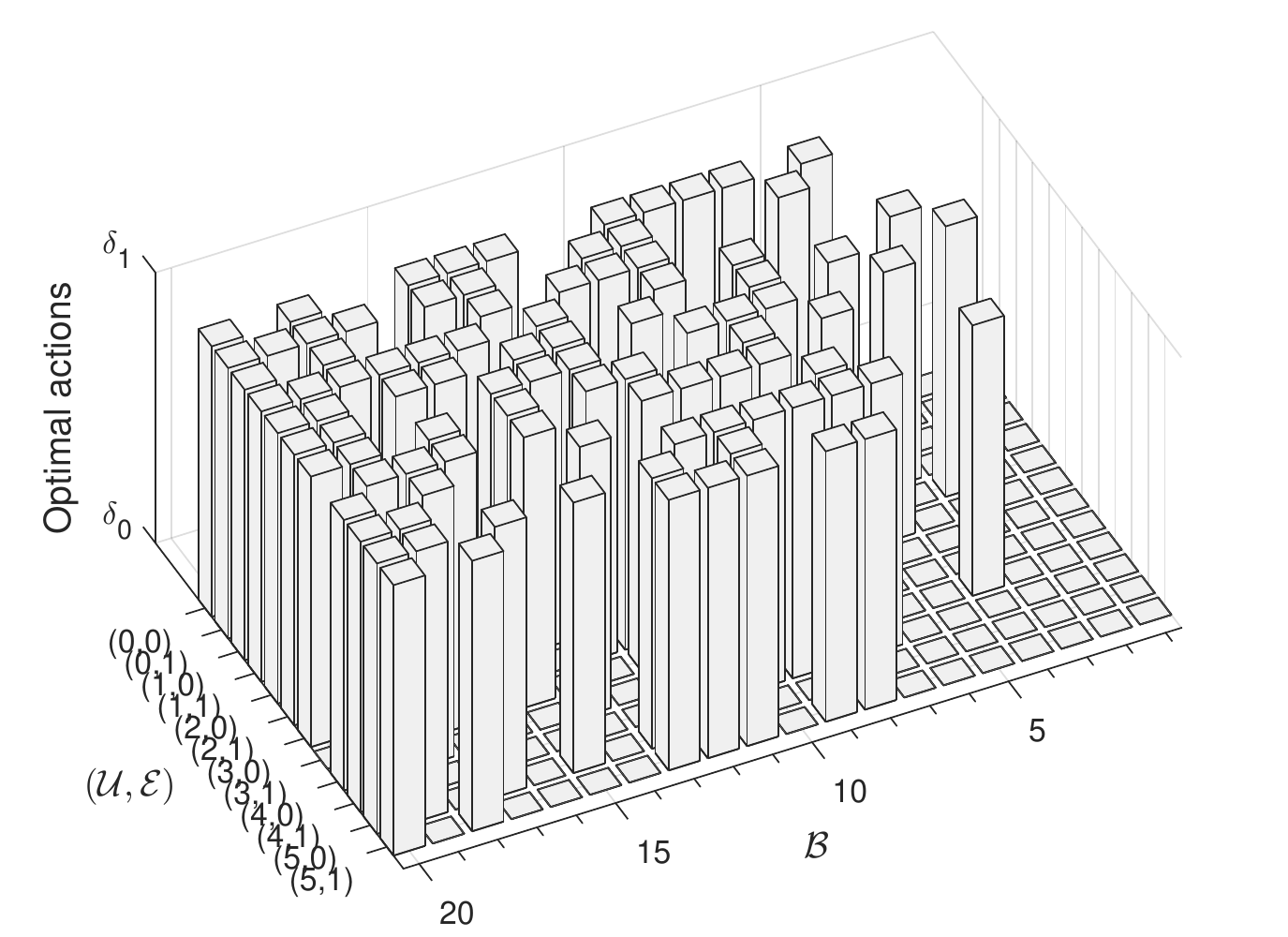}
\par\end{centering}

}\subfloat[\label{fig:structured_q_learning_2}]{\begin{centering}
\includegraphics[width=0.9\columnwidth]{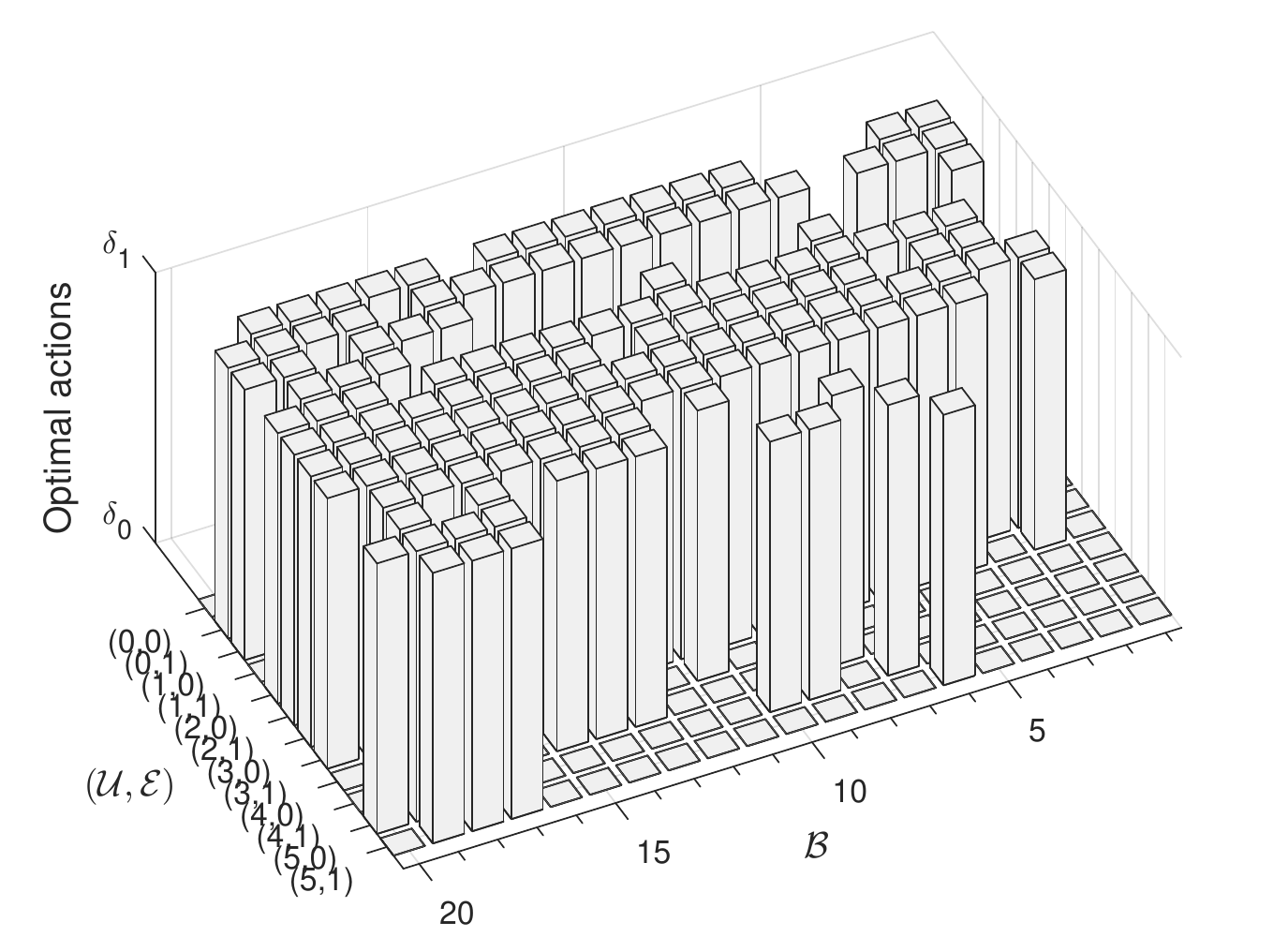}
\par\end{centering}

}
\par\end{centering}

\caption{Online learning of the optimal activity tracking policy using $10^{6}$ gradual iterations. (a)~The poor estimation using conventional Q-learning, and (b)~a better estimation using the structured Q-learning algorithm.\label{fig:structured_q_learning}}

\end{figure*}

The key objective of proving the monotone threshold structure of the activity tracking policy in a mobile device is for adopting low-complexity estimation methods as discussed in Section~\ref{sub:structured_q_learning}. Figure~\ref{fig:structured_q_learning} shows the online policy learning by applying the Q-learning algorithm. In particular, Figure~\ref{fig:structured_q_learning_1} shows the policy which is generated after $10^{6}$ update iterations of the conventional Q-learning method. Clearly, the conventional Q-learning does not consider the monotone structure of the policy and it initializes the state-action function to random or zero values. Then, it searches through the whole solution space, i.e.,  a brute-force search. This causes the poor performance of the conventional method. By contrast, the structured Q-learning algorithm starts with an initialization that considers the actual monotonicity of the state-action function and projects the final policy into a threshold form. Consequently, this significantly improves the performance as shown in Figure~\ref{fig:structured_q_learning_2}.

\subsection{Performance Metrics}

We consider three performance metrics of online activity tracking systems under a data usage limit.

\subsubsection{Average Battery Level}

This measure is important to ensure that the mobile device has enough energy when performing the activity tracking task. The average battery level $\bar{b}$ during the active mode $\delta_{1}$ can be obtained as follows:
\begin{equation}
\bar{b}=\sum\limits _{u\in\mathcal{U}}\sum\limits _{e\in\mathcal{E}}\sum\limits _{b\in\mathcal{B}}b	\phi^{*}(\psi=[u,e,b],\delta_{1}).
\end{equation}

\subsubsection{Probability of Successful Data Synchronization}

Successful data synchronization to a backend server requires (i)~the selection of the activation action, i.e., $\delta_{1}$, and (ii)~the availability of access network connection determined by $g(\psi,\delta)$. Therefore, we define the probability of successful data synchronization $\rho$ as follows:
\begin{equation}
\rho=\sum\limits _{u\in\mathcal{U}}\sum\limits _{e\in\mathcal{E}}\sum\limits _{b\in\mathcal{B}}g(\psi,\delta_{1})	\phi^{*}(\psi=[u,e,b],\delta_{1}).
\end{equation}

\subsubsection{Probability of Battery Overflow}

Assuming that the mobile device is only used for activity tracking, this metric measures the probability of overcharging the battery of the mobile device by adding an energy unit to a fully charged battery, i.e., the probability of wasting energy. This can happen in two cases: (i)~charging a full battery during the sleep mode, or (ii)~charging a full battery when no access network coverage is available during the active mode. Then, the probability of battery overflow $\tau$ is given by
\begin{multline}
\tau=\sum\limits _{u\in\mathcal{U}}\phi^{*}(\psi=[u,1,B-1],\delta_{1})\left[1-g(\psi,\delta_{1})\right]\\
+\phi^{*}(\psi=[u,1,B-1],\delta_{0}).
\end{multline}

\subsection{Performance Evaluations}

In this section, the optimal CMDP policy is compared with a baseline activity tracking policy. We consider the baseline policy that also guarantees monetary access cost through a cellular data limit $D$, and hence it achieves the data transmission probability $\xi$ of the total epochs as in (\ref{eq:coverage_percentage}). We propose a \emph{constrained uniform policy (CUP)} as a baseline method, such that the mobile device is activated based on a fixed stationary probability and $\phi_{\textrm{CUP}}^{*}(\psi,\delta)$ is uniform for all states $\psi\in\Psi$ and action $\delta\in\Delta$. Recall that the policy should sample the user activity with a probability $\xi$ of its total decision epochs. Thereby, the uniform probability is simply found as follows:
\begin{equation}
\phi_{\textrm{CUP}}^{*}(\psi,\delta)=\begin{cases}
\frac{1-\xi}{U\times B\times2}, & \delta=\delta_{0},\\
\frac{\xi}{U\times B\times2}, & \delta=\delta_{1}.
\end{cases}
\end{equation}
where the constraints $\sum\limits _{\psi\in\Psi_{\textrm{CUP}}}\sum\limits _{\delta\in\Delta}\phi_{\textrm{CUP}}(\psi,\delta)=1$ and $\phi(\psi,\delta)\geq0$ are satisfied. In the following, we vary the system parameters and observe their impact on the performance of the data usage-constrained policies.

\subsubsection{Detection Error}
\begin{figure}
\begin{centering}
\includegraphics[width=0.85\columnwidth]{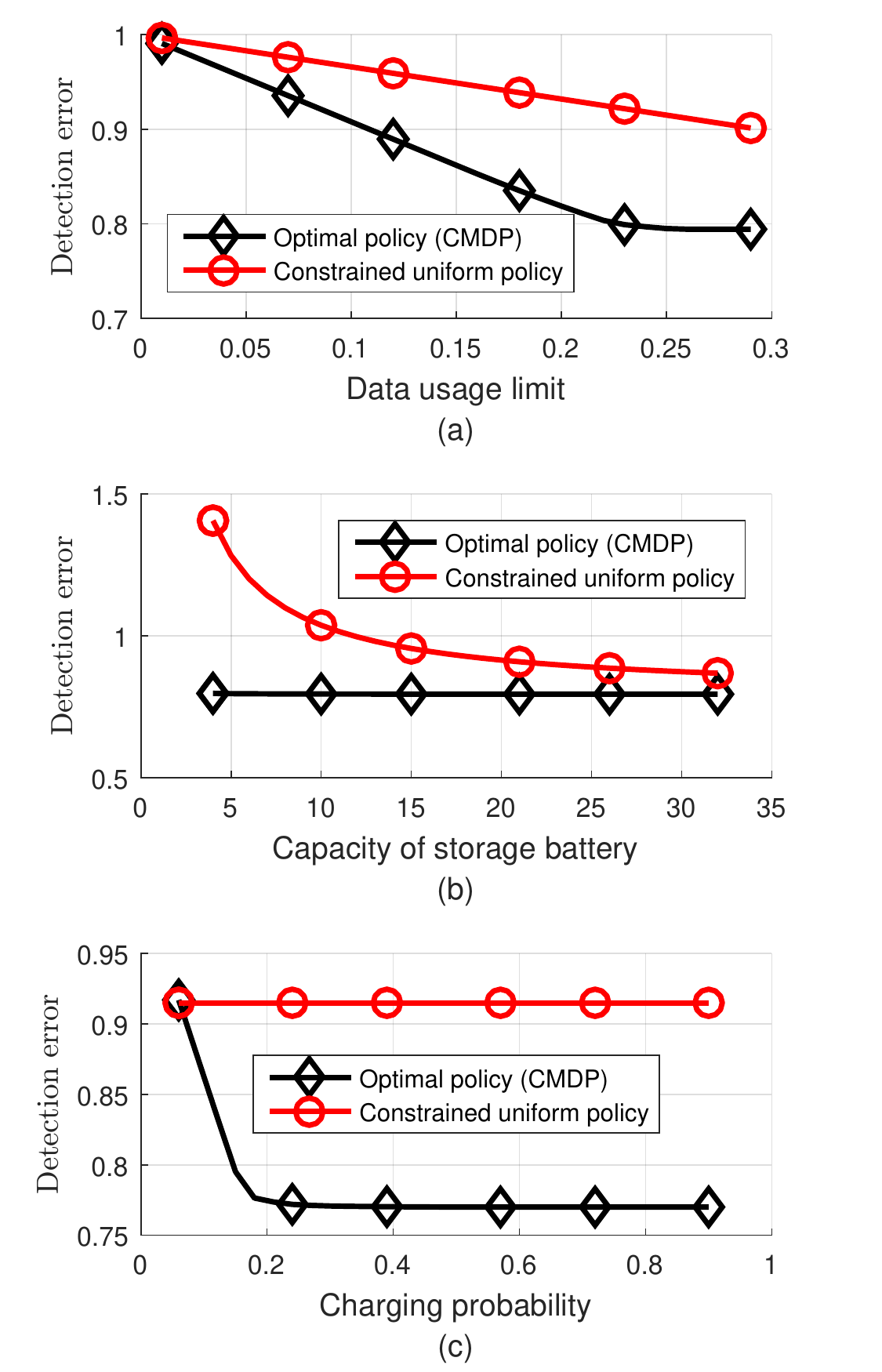}
\par\end{centering}
\caption{Detection error $\mathcal{J}\left(\pi\right)$ under varied data usage limit \textbf{$D$}, capacity of storage battery $B$, and charging probability $\mathbb{P}(e=1)$.\label{fig:detection_err}}
\end{figure}

Figure~\ref{fig:detection_err} shows the performance of the optimal CMDP policy and the constrained uniform policy when the data usage limit $D$, capacity of storage battery $B$, and charging probability $\mathbb{P}(e=1)$ are varied. Several important results can be observed. Firstly, as $D$ becomes more relaxed, the optimal policy senses the user activity more frequently which decreases the detecting error $\mathcal{J}\left(\pi\right)$. This indicates that if a user decides to set a low data usage setup, the system tracking of that particular user will be poor. Secondly, the detection error will be slightly decreased when the capacity of storage battery $B$ is increased. This is intuitive as the extra battery storage helps in decreasing the battery overflow during the charging process. Thirdly, when the charging probability is high, the detection error is low due to the increased energy budget of the mobile device. It is important to note that the charging probability depends on the number of charger deployed in the movement locations. It can be observed that the optimal policy outperforms the constrained uniform policy in all scenarios.

\begin{figure}
\begin{centering}
\includegraphics[width=0.85\columnwidth,trim=0cm 0cm 0cm 0cm]{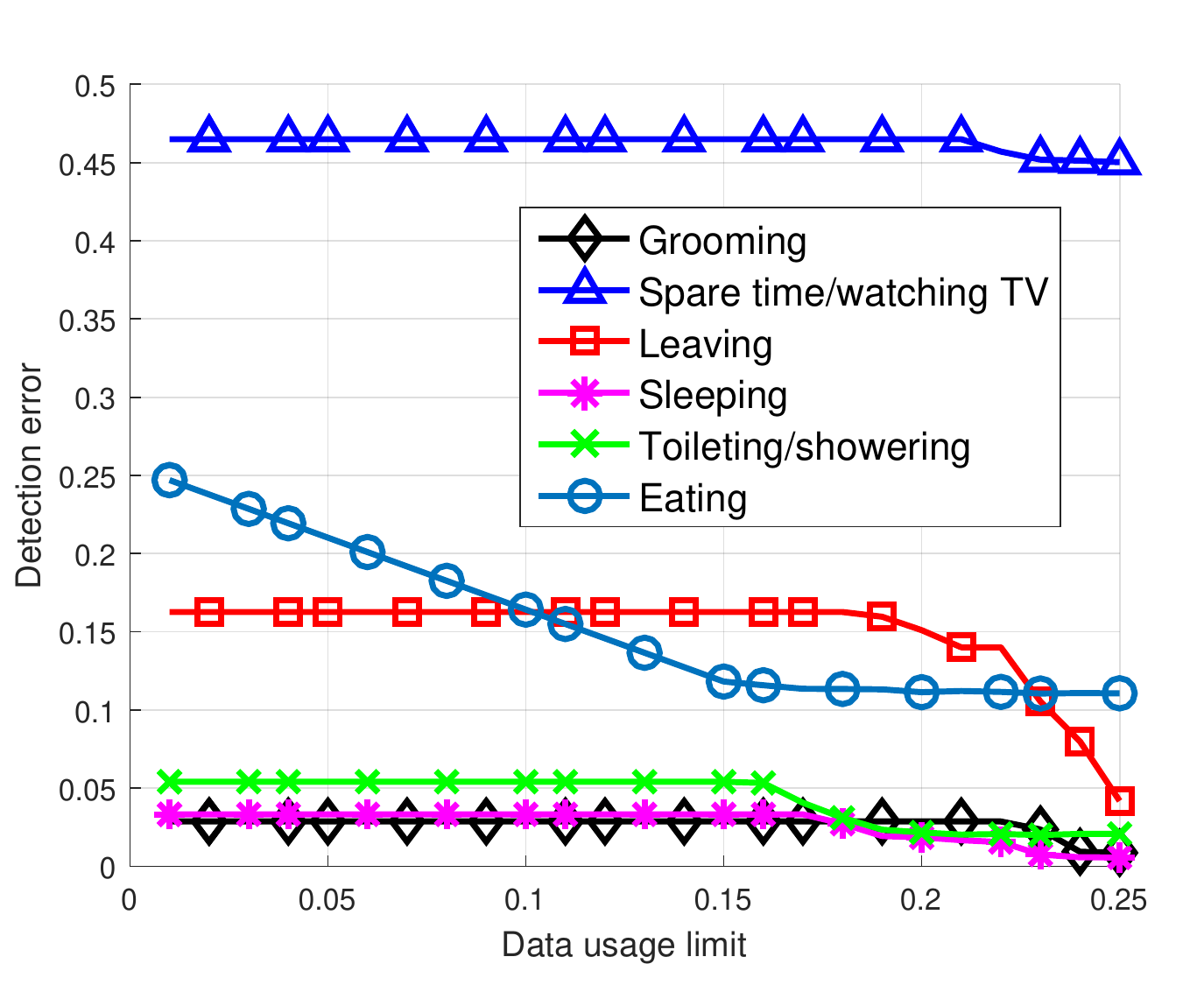}
\par\end{centering}
\caption{Detection error per activity of the optimal tracking policy under varied data usage limit \textbf{$D$}.\label{fig:err_per_activity}}
\end{figure}

Figure~\ref{fig:err_per_activity} shows the average detection error for each activity under varied data usage limit $D$. It can be noted that the detection error of each activity decreases as the data usage limit $D$ is increased. This is due to the increased rate of synchronization probability as defined in (\ref{eq:coverage_percentage}). However, it can be noted that the average detection error does not uniformly decrease for all activities. Instead, the optimal policy defines the optimal activation based on the detection error and transition probabilities of different activities with the objective of minimizing the total detection error of the tracking system. 

\subsubsection{Maximum Capacity of Battery}
\begin{figure}
\begin{centering}
\includegraphics[width=0.85\columnwidth,trim=0cm 0.5cm 0cm 0cm]{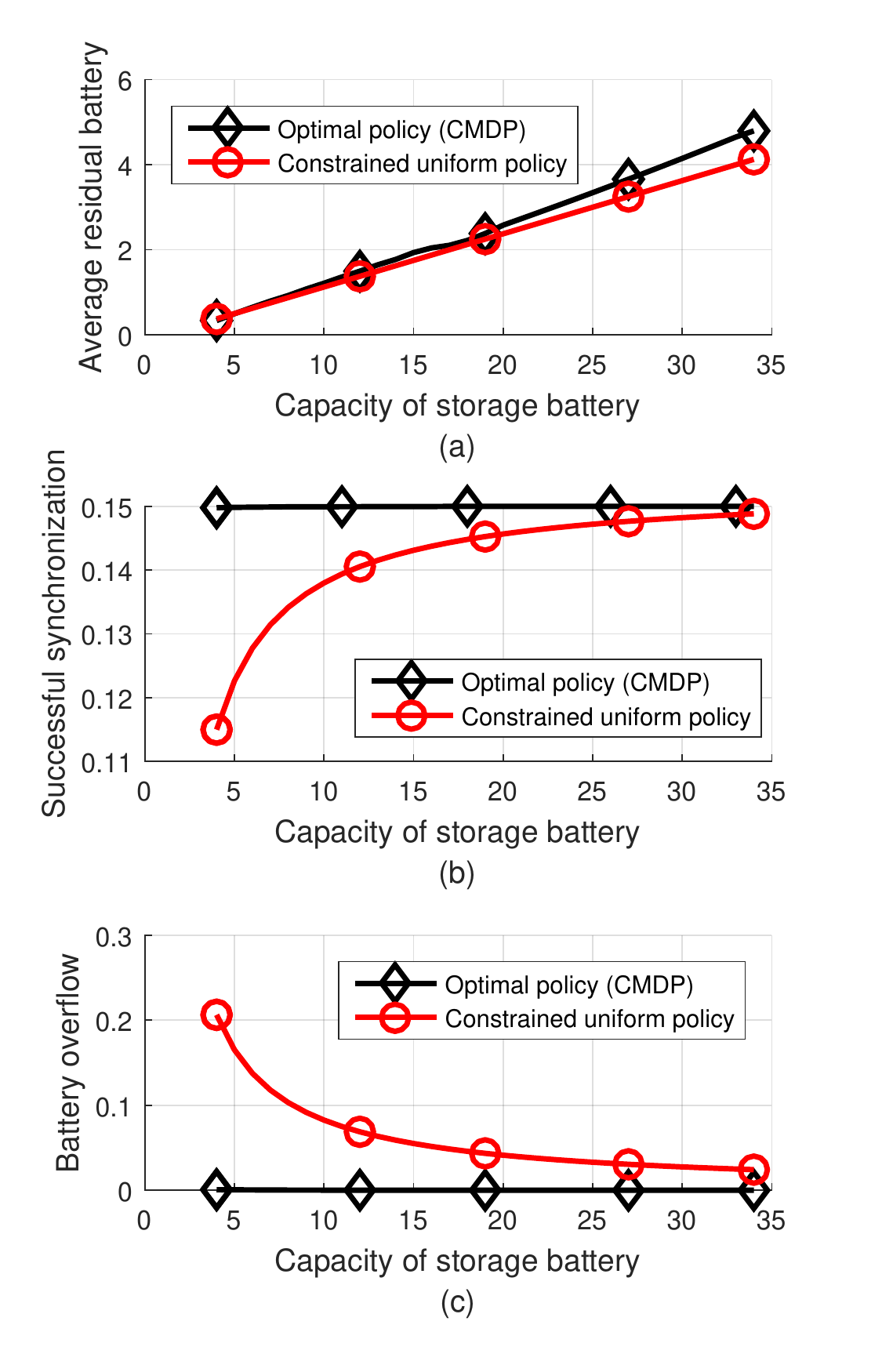}
\par\end{centering}

\caption{Average battery level $\bar{b}$ (in energy units), probability of successful data synchronization $\rho$, and probability of battery overflow $\tau$ under varied maximum capacities of battery storage $B$ (in energy units). The data usage limit $D$ is fixed at $0.15$.\label{fig:battery_storage_capacity}}
\end{figure}

Figure~\ref{fig:battery_storage_capacity} shows the performance of optimal CMDP policy and the constrained uniform policy when the maximum capacity of storage battery $B$ is varied. When $B$ is high, the average battery level $\bar{b}$ increases as the battery can store more energy units. Likewise, the probability of successful data synchronization $\rho$ slightly increases. The probability of battery overflow $\tau$ slightly decreases. The optimal policy outperforms the constrained uniform policy in all performance metrics.

%
%
%

\section{Conclusions and Future Work}\label{sec:conclusions}

In this paper, we have presented an activity tracking policy with its threshold structure for mobile devices with intermittent energy charging. We have first modeled the activity and motion tracking as a stochastic optimization using constrained Markov decision processes (CMDPs). The objective is to minimize the detection error of human activities subject to a data usage limit. The CMDP-based problem has been then transformed into an unconstrained, discounted cost MDP with infinite time horizon by using a Lagrange relaxation method. Specifically, a Lagrange multiplier is used to capture the cellular data limit, and therefore ensures the monetary access requirement in the unconstrained activity tracking policy. The CMDP policy has been proved to be a randomized mixture of two threshold MDP policies. Equally important, the CMDP policy has been shown to be monotonically non-decreasing in the battery level. This monotone threshold structure enables a fast and online learning of the optimal activity tracking policy when the user statistics are unknown a priori, time-varying, and user-defined.

For the future work, dynamic energy pricing can be introduced for wireless charging and the activity tracking policy has to optimize this cost.


\section*{Appendix}
\subsection{Proof of Lemma~\ref{lem:value_monotonicity}}
The monotonic structure of $v(\psi;\lambda,\beta)$ can be proved by showing that the two terms of (\ref{eq:bellman_equation}) are monotone in the system states. In other words, by proving that (a)~the Lagrangian error function is monotone, and (b)~the transition probability summation is also monotone. Firstly, $c(\psi,\delta;\lambda)$ is monotonically decreasing in the user activity $u$ and monotonically non-decreasing in battery level $b$. This is clear from the definitions of the cost and data usage functions in (\ref{eq:cost_function}) and (\ref{eq:data_usage_function}), respectively. Secondly, the transition probability summation $\sum\limits _{\psi'\in\Psi}\mathbb{P}(\psi'|\psi,\delta)$ is also monotonically non-decreasing in the battery level $b$ as the transition probabilities are assumed to satisfy the first-order stochastic dominance rule.

\subsection{Proof of Theorem~\ref{thm:monotone_policy}}
The optimal discounted cost MDP policy $\pi^{*}_{\text{MDP}}$ can be shown to be monotone by inductively proving that the state-action cost function $Q(\psi,\delta;\lambda,\beta)$, calculated using (\ref{eq:state_action_cost}) for $i\in\left\{ 0,1,2,\ldots\right\} $, is a submodular function in $(b,\delta)$. Mathematically, this can by proved by showing that
\begin{multline}
Q^{i+1}(\psi=\left[u,e,b\right],\delta_{1};\lambda,\beta)\\
-Q^{i+1}(\psi=\left[u,e,b\right],\delta_{0};\lambda,\beta)\geq\\
Q^{i+1}(\psi=\left[u,e,b+1\right],\delta_{1};\lambda,\beta)\\
-Q^{i+1}(\psi=\left[u,e,b+1\right],\delta_{0};\lambda,\beta),
\label{eq:q_function_monotonicity}
\end{multline}
which indicates that $Q^{i+1}(\psi,\delta;\lambda,\beta)$ is submodular in $(b,\delta)$ for all $i\in\left\{ 0,1,2,\ldots\right\} $. Using (\ref{eq:state_action_cost}) and (\ref{eq:transition}), the left hand side (LHS) of (\ref{eq:q_function_monotonicity}) can be rewritten as follows:
\begin{multline}
Q^{i+1}\left(\psi=\left[u,e,b\right],\delta_{1};\lambda,\beta\right)\\
-Q^{i+1}\left(\psi=\left[u,e,b\right],\delta_{0};\lambda,\beta\right)\\
=c\left(\psi=\left[u,e,b\right],\delta_{1};\lambda\right)-c\left(\psi=\left[u,e,b\right],\delta_{0};\lambda\right)\\
+\beta\sum\limits _{\psi'\in\Psi}\mathbb{P}(u'|u)\mathbb{P}(e')g(\psi,\delta_{1})\\
\times\left[v^{i}\left(\psi'=\left[u',e',b'-1\right];\lambda,\beta\right)-v^{i}\left(\psi'=\left[u',e',b'\right];\lambda,\beta\right)\right],
\label{eq:q_function_expand}
\end{multline}
where $c\left(\psi=\left[u,e,b\right],\delta_{0};\lambda\right)$ is equal to zero by the problem setup. Then, the inequality condition in (\ref{eq:q_function_monotonicity}) can be proved by showing that $v^{i}\left(\psi;\lambda,\beta\right)$ is non-decreasing in $b$ for all $i\in\left\{ 0,1,2,\ldots\right\} $ such that
\begin{equation}
\begin{array}{cc}
v^{i+1}(\left[u,e,b+1\right];\lambda,\beta)-v^{i+1}(\left[u,e,b\right];\lambda,\beta)\geq\\
v^{i+1}(\left[u,e,b\right];\lambda,\beta)-v^{i+1}(\left[u,e,b-1\right];\lambda,\beta),
\end{array}\label{eq:value_function_monotonicity}
\end{equation}
or equivalently
\begin{equation}
\begin{array}{cc}
v^{i+1}(\left[u,e,b+1\right];\lambda,\beta)-v^{i+1}(\left[u,e,b\right];\lambda,\beta)\\
-\left[v^{i+1}(\left[u,e,b\right];\lambda,\beta)-v^{i+1}(\left[u,e,b-1\right];\lambda,\beta)\right] & \geq0.
\end{array}\label{eq:value_function_monotonicity_1}
\end{equation}
Recall that $v\left(\psi;\lambda,\beta\right)$ is defined for states, and $Q(\psi,\delta;\lambda,\beta)$ is defined for state-action pairs. For actions $\delta^{0},\delta^{1},\delta^{2}\in\Delta$ which are selected such that
\begin{eqnarray}
v^{i+1}(\left[u,e,b-1\right];\lambda,\beta)= Q^{i+1}(\left[u,e,b-1\right],\delta^{0};\lambda,\beta),\label{eq:optimality_assumption_1}\\
v^{i+1}(\left[u,e,b\right];\lambda,\beta) = Q^{i+1}(\left[u,e,b\right],\delta^{1};\lambda,\beta),\textrm{ and}\label{eq:optimality_assumption_2}\\
v^{i+1}(\left[u,e,b+1\right];\lambda,\beta)=Q^{i+1}(\left[u,e,b+1\right],\delta^{2};\lambda,\beta),\label{eq:optimality_assumption_3}
\end{eqnarray}
which means that $\delta^{0},\delta^{1}$, and $\delta^{2}$ are optimal actions at states $\left[u,e,b-1\right],\left[u,e,b\right]$, and $\left[u,e,b+1\right]$, respectively. Then, the right hand side (RHS) of (\ref{eq:value_function_monotonicity_1}) can be expressed as follows:
\begin{equation}
\begin{array}{cc}
Q^{i+1}(\left[u,e,b+1\right],\delta^{2};\lambda,\beta)-Q^{i+1}(\left[u,e,b\right],\delta^{1};\lambda,\beta)\\
-Q^{i+1}(\left[u,e,b\right],\delta^{1};\lambda,\beta)+Q^{i+1}(\left[u,e,b-1\right],\delta^{0};\lambda,\beta).
\end{array}\label{eq:value_function_monotonicity_2}
\end{equation}
By adding $Q^{i+1}(\left[u,e,b\right],\delta^{2};\lambda,\beta)$ and $Q^{i+1}(\left[u,e,b\right],\delta^{0};\lambda,\beta)$ with their negative values, the expression in~(\ref{eq:value_function_monotonicity_2}) becomes
\begin{equation}
\begin{array}{cc}
\underset{\text{Term 1}}{\underbrace{Q^{i+1}(\left[u,e,b+1\right],\delta^{2};\lambda,\beta)-Q^{i+1}(\left[u,e,b\right],\delta^{2};\lambda,\beta)}}\\
+\underset{\text{Term 2}}{\underbrace{Q^{i+1}(\left[u,e,b\right],\delta^{2};\lambda,\beta)-Q^{i+1}(\left[u,e,b\right],\delta^{1};\lambda,\beta)}}\\
\underset{\text{Term 3}}{+\underbrace{Q^{i+1}(\left[u,e,b\right],\delta^{0};\lambda,\beta)-Q^{i+1}(\left[u,e,b\right],\delta^{1};\lambda,\beta)}}\\
-\underset{\text{Term 4}}{\underbrace{\left[Q^{i+1}(\left[u,e,b\right],\delta^{0};\lambda,\beta)-Q^{i+1}(\left[u,e,b-1\right],\delta^{0};\lambda,\beta)\right]}.}
\end{array}\label{eq:value_function_monotonicity_3}
\end{equation}
Terms~2 and~3 are positive in magnitude by the assumptions given in (\ref{eq:optimality_assumption_1})-(\ref{eq:optimality_assumption_3}) of optimal actions. In particular, $\delta^{1}$ is defined as an optimal action at state $\left[u,e,b\right]$, and hence it has the minimum state-action value. Moreover, Term~4 is less than Term~1, which can be shown by expanding these terms as in (\ref{eq:q_function_expand}) and knowing that $v(\psi;\lambda,\beta)$ is monotonically non-decreasing in the battery level $b$. This proves that the condition in (\ref{eq:value_function_monotonicity_1}) is satisfied, and hence $Q^{i+1}(\psi,\delta;\lambda,\beta)$ is submodular in $(b,\delta)$ for all $i\in\left\{ 0,1,2,\ldots\right\} $. This proves that the optimal discounted cost MDP policy $\pi_{\text{MDP}}^{*}$ is also monotonically non-decreasing in $(b,\delta)$.

\section*{Acknowledgment}
This work was supported in part by the National Research Foundation of Korea (NRF) grant funded by the Korean government (MSIP) (2014R1A5A1011478), Singapore MOE Tier~1 under Grant RG122/15 and Grant RG18/13, and Singapore MOE Tier~2 under Grant MOE2013-T2-2-070 ARC16/14 and Grant MOE2014-T2-2-015 ARC 4/15. We thank Zhang~Yang for his valuable comments in the early stages of this work.

\bibliographystyle{IEEEtran}
\bibliography{mobile_sensing}

\begin{IEEEbiography}[{\includegraphics[width=1in,height=1.25in,clip,keepaspectratio]{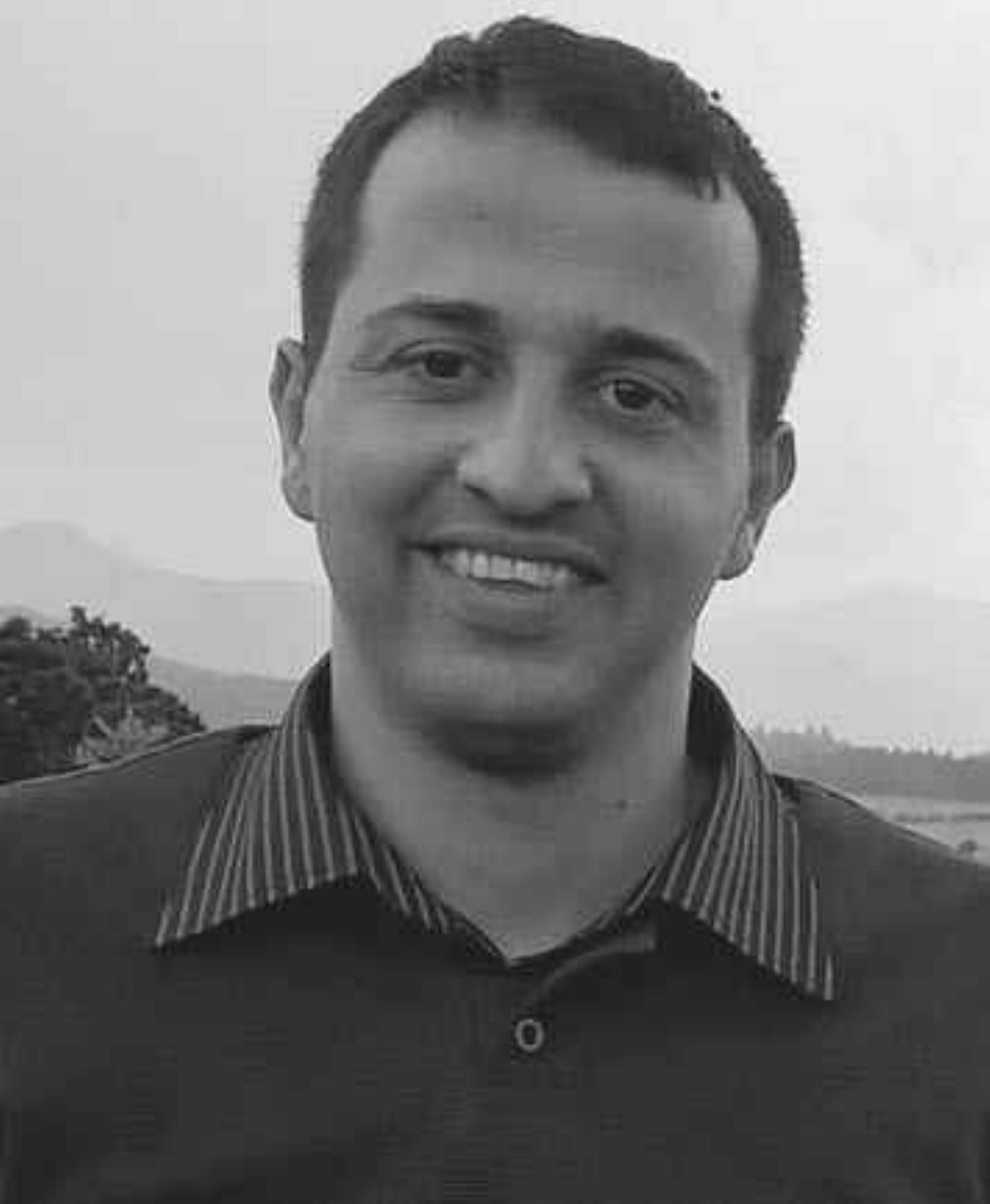}}]
{Mohammad Abu Alsheikh}(S'14)
received his B.Eng. in computer systems engineering from Birzeit University, Palestine, in 2011. Between 2010 and 2012, he was a Software Engineer working on developing robust web services, Ajax-based web components, and smartphone applications. He is currently a Ph.D. candidate in the School of Computer Engineering, Nanyang Technological University, Singapore. His research interests include machine learning in big data analytics, mobile sensing technologies, and sensor-based activity recognition.
\end{IEEEbiography}

\begin{IEEEbiography}[{\includegraphics[width=1in,height=1.25in,clip,keepaspectratio]{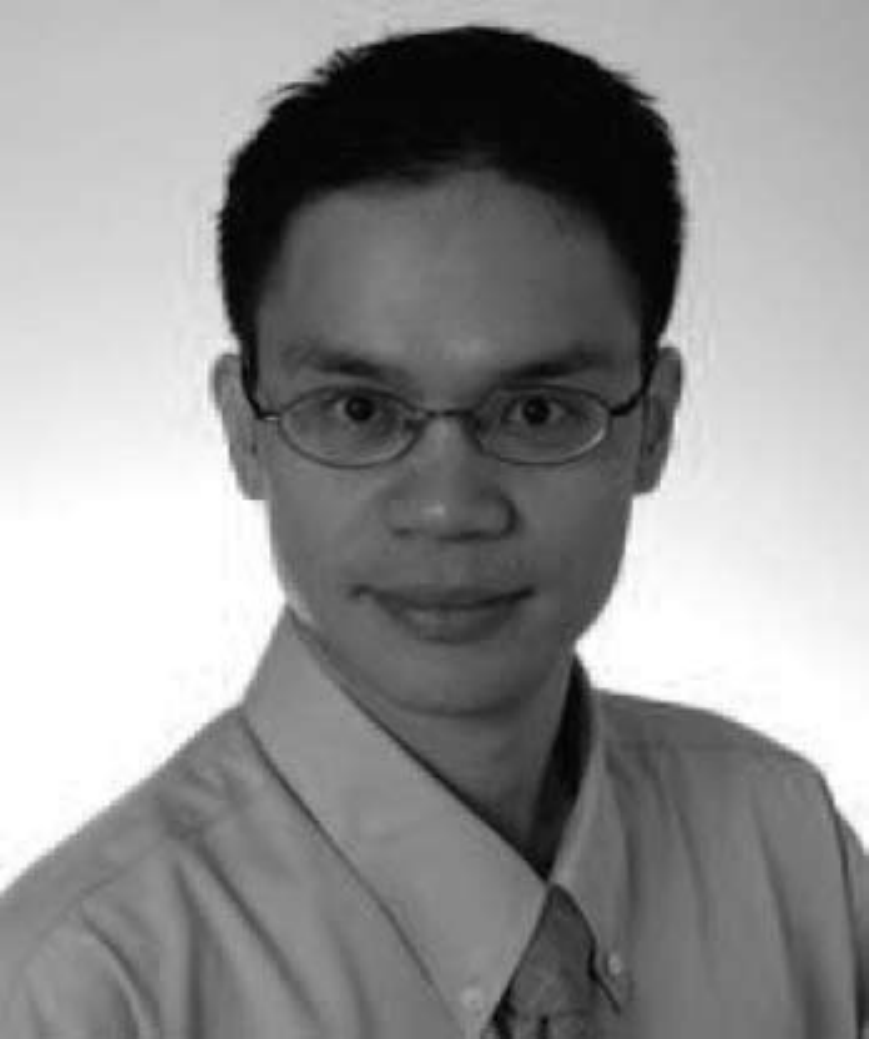}}]
{Dusit Niyato}(M'09--SM'15)
is currently an Associate Professor in the School of Computer Science and Engineering, at Nanyang Technological University, Singapore. He received B.Eng. from King Mongkuts Institute of Technology Ladkrabang (KMITL), Thailand in 1999 and Ph.D. in Electrical and Computer Engineering from the University of Manitoba, Canada in 2008. His research interests are in the area of energy harvesting for wireless communication, Internet of Things (IoT) and sensor networks.
\end{IEEEbiography}
\vfill

\begin{IEEEbiography}[{\includegraphics[width=1in,height=1.25in,clip,keepaspectratio]{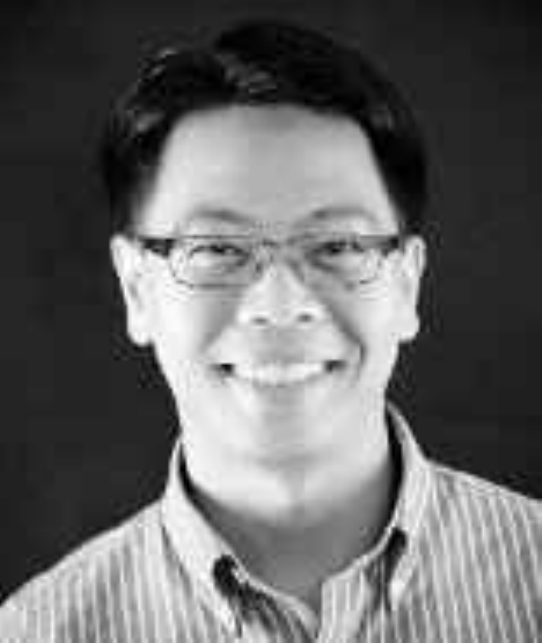}}]
{Shaowei Lin}
received his Ph.D. in Mathematics under Bernd Sturmfels in 2011 from the University of California, Berkeley, where he analyzed singularities in statistical models over large data sets through the lens of modern algebraic geometry. This work was continued at Stanford University in a one-year DARPA postdoctoral collaboration with Andrew Ng's lab to explore mathematical challenges in deep learning. In 2012, he returned to Singapore to join the Institute for Infocomm Research (A*STAR) where he started the Sense-making Group in the Sense and Sense-abilities (S\&S) programme. The group focused on exploiting machine learning techniques in sensor networks to create resource-efficient algorithms that exhibit higher-order intelligence. Before joining Singapore University of Technology and Design (SUTD), he oversaw deep science activities in S\&S as the Deputy Head for Research.
\end{IEEEbiography}

\begin{IEEEbiography}[{\includegraphics[width=1in,height=1.25in,clip,keepaspectratio]{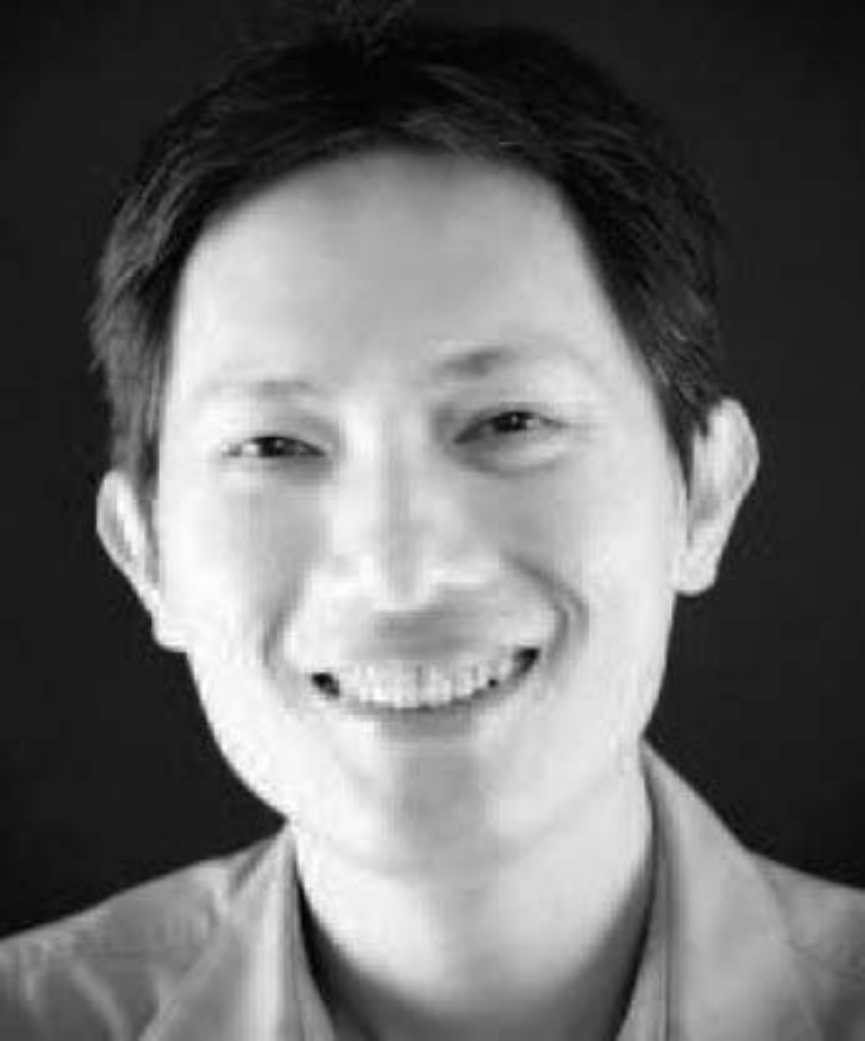}}]
{Hwee-Pink TAN}(S'00--M'04--SM'14)
is currently an Associate Professor of Information Systems (Practice) as well as the Academic Director of the TCS-SMU iCity Lab at the Singapore Management University. Prior to joining SMU in March 2015, he was the SERC Programme Manager of the A*STAR Sense and Sense-abilities Program, Institute for Infocomm Research (I\textsuperscript{2}R), Singapore. Before returning to Singapore to join I\textsuperscript{2}R in March 2008, he was a Research Fellow at Center for Telecommunications Value-chain Research (CTVR) in the Emerging Networks (EN) strand, led by Dr. Linda Doyle. Between December 2004 and June 2006, he was a post-doctoral researcher at EURANDOM in the research group, Queueing and Performance Analysis (QPA), under the guidance of Prof. Onno Boxma and Dr. Ivo Adan. He also visited the Telecommunications Group at the University of Ferrara from September-October 2004 and was hosted by Prof. Michele Zorzi, who is a Professor of Telecommunications in the Department of Information Engineering, University of Padua, Italy.
\end{IEEEbiography}

\begin{IEEEbiography}[{\includegraphics[width=1in,height=1.25in,clip,keepaspectratio]{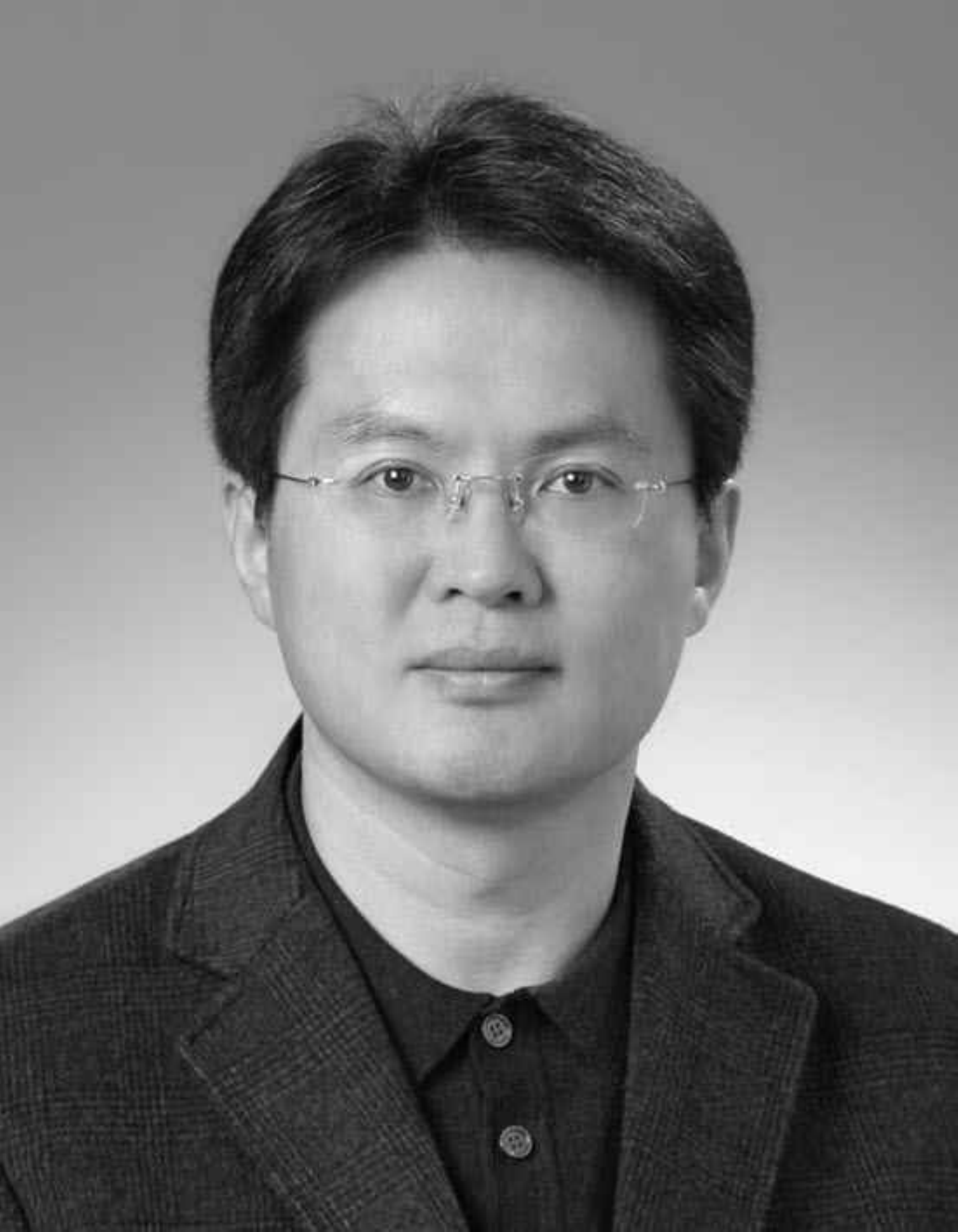}}]
{Dong In Kim}(S'89-M'91-SM'02)
received the Ph.D. degree in electrical engineering from the University of Southern California, Los Angeles, CA, USA, in 1990. He was a Tenured Professor with the School of Engineering Science, Simon Fraser University, Burnaby, BC, Canada. Since 2007, he has been with Sungkyunkwan University (SKKU), Suwon, Korea, where he is currently a Professor with the College of Information and Communication Engineering. Dr. Kim is a first recipient of the NRF of Korea Engineering Research Center in Wireless Communications for Energy Harvesting Wireless Communications (2014-2021). From 2002 to 2011, he served as an Editor and a Founding Area Editor of Cross-Layer Design and Optimization for the IEEE Transactions on Wireless Communications. From 2008 to 2011, he served as the Co-Editor-in-Chief for the IEEE/KICS Journal of Communications and Networks. He served as the Founding Editor-in-Chief for the IEEE Wireless Communications Letters from 2012 to 2015. From 2001 to 2014, he served as an Editor of Spread Spectrum Transmission and Access for the IEEE Transactions on Communications, and then serving as an Editor-at-Large of Wireless Communication I for the IEEE Transactions on Communications. 
\end{IEEEbiography}
\vfill

\end{document}